\input{psfig}
\documentclass[]{aa}
\usepackage{graphicx}
\usepackage{deluxetable}
\usepackage{aalongtable}
\begin{document}
\title{RACE-OC Project:\\ Rotation and variability in the open cluster NGC\,2099 
(M37)\thanks{Tables 1-2 and Figs.\,11-26 are available
 only in electronic form.}}
\author{S.\,Messina\inst{1}
\and          E.\,Distefano\inst{1}
\and          Padmakar Parihar\inst{2}
\and          Y.\,B.\,Kang\inst{3,4,5}
\and          S.\,-L.\,Kim\inst{4}
\and          S.\,-C.\,Rey\inst{3}
\and          C.\,-U.\,Lee\inst{4}}
\offprints{Sergio Messina}
\institute{INAF-Catania Astrophysical Observatory, via S.\,Sofia 78, I-95123 Catania, Italy  \\
\email{sergio.messina@oact.inaf.it; elisa.distefano@oact.inaf.it}
\and   
Indian Institute of Astrophysics, Block II, Koramangala, Bangalore India, 560034 \\
\email{psp@iiap.res.in}
\and
Department of Astronomy and Space Science, Chungnam National University, Daejeon, Korea
\and
Korea Astronomy and Space Science Institute,  Daejeon, Korea
\and
Department of Physics and Astronomy, Johns Hopkins University,
Baltimore, MD 21218, USA\\
}

\date{}
\titlerunning{Rotation and variability in NGC\,2099}
\authorrunning{S.\,Messina et al.}
\abstract {Rotation and solar-type magnetic activity are closely related to each other in   main-sequence stars of G or later spectral 
types. Presence and level of magnetic activity depend on star's rotation and rotation  itself is strongly influenced by 
strength and topology of  the magnetic fields. Open clusters represent  especially useful  targets to investigate the connection 
between rotation and activity.}
 { The open cluster NGC\,2099 has been studied as a part of the RACE-OC project, which is aimed at exploring the evolution 
of rotation and magnetic activity in the late-type members of open clusters of different ages.} {Time series CCD 
photometric observations of this cluster were collected during January 2004  and the presence of periodicities in the 
flux variation related to the stellar  rotation are determined by Fourier analysis. The relations between activity 
manifestations, such as the light curve amplitude, and global stellar parameters are investigated.} {We have discovered 
135 periodic variables, 122 of which are candidate cluster members. Determination of rotation periods of G- and K-type 
stars has  allowed us  to better explore evolution of angular momentum   at an age of about  500 Myr.  In our analysis  
we have also identified  3  new detached eclipsing binary candidates among cluster  members. } {A comparison with the older
Hyades  cluster ($\sim$625 Myr) shows that the newly determined distribution of rotation periods 
is consistent with the scenario of   rotational braking  of main-sequence spotted stars as they age.  However, 
a comparison
with the younger M34 cluster ($\sim$200 Myr) shows that the G8-K5 members of these clusters have the same rotation period distribution, 
that is G8-K5 members in NGC\,2099 seem to have experienced no significant  braking in the age range from $\sim$200 to $\sim$500 Myr.
Finally, NGC\,2099 members have a level of photospheric magnetic activity, as measured by light curve amplitude, smaller than in younger stars of same mass and rotation, suggesting that the activity level also depends on some other age-dependent parameters. }
\keywords{Stars: activity - Stars: binaries: eclipsing - Stars: late-type - Stars: rotation - 
Stars: starspots - Stars: open clusters and associations: individual: NGC\,2099}
\maketitle
\rm
\section{Introduction}

The RACE-OC project, which stands for {\bf R}otation and {\bf AC}tivity {\bf E}volution in {\bf O}pen {\bf C}lusters, is a long-term project aimed at
 studying the evolution  of the rotational properties and the magnetic activity of late-type members of stellar open
  clusters (Messina 2007a).
Stellar open clusters represent privileged astrophysical targets since they provide complete and homogeneous stellar
 samples to explore a variety of relevant problems of  astrophysical impact. The  homogeneity of the stellar sample arises from the fact that
all cluster members were formed in similar environmental conditions, characterized by same age as well as initial chemical composition, and are subjected to same interstellar reddening. Such  complete stellar samples  allow us to accurately investigate those stellar properties, and their mutual relations, which depend on age and metallicity.\\
\indent
Indeed, rotation is one of  the basic stellar properties on which the present project is focused. It undergoes dramatic changes along the whole stellar life, as predicted by evolution models  of angular momentum of late-type stars (Kawaler 1988; MacGregor \& Brenner 1991; Krisnamurthi et al. 1997; Bouvier et al. 1997; Sills et al. 2000; Ivanova \& Taam 2003; Holzwarth \& Jardine 2007). However, a growing body of  evidence, especially from
 observational studies of open clusters, shows that a significant discrepancy exists with respect to the theoretical expected
  scenario. The observed spread of rotation periods among stars of similar mass, age and chemical composition (see e.g. Rebull et al. 2004), the existence of slowly rotating PMS stars (Herbst \& Mundt 2005), and the well-known decay of rotation rate shown by main-sequence stars when they age at approximately constant radius (see e.g. Barnes 2007), are just a few examples of the
 mentioned discrepancy with respect to predictions of evolutionary models. \\
\indent 
It is believed that strong magnetic fields play a
  fundamental role in altering the rotational properties of late-type stars. They are responsible for angular momentum loss or its internal redistribution, and represent a powerful tool to probe the stellar internal structure. For example, 
 magnetic fields are believed to play a key role in the distribution of the mass moment of inertia by coupling
 the radiative core with the external convection zone (e.g. Barnes 2003). The study of angular momentum evolution is very important to better understand the magnetic activity phenomena which
 manifest themselves in late-type stars and directly depend on  stellar rotation. Photospheric cool spots and bright faculae,
  chromospheric plages and X-ray emission all arise from the action of an hydromagnetic dynamo whose efficiency is related
   to the star's rotation rate (e.g. Messina et al. 2001; 2003). \\
 \indent 
 Thus, the study of evolution of angular momentum and magnetic activity offers a complementary approach to understand the
 mechanisms by which rotation and magnetic fields influence each other and, eventually, to better understand the nature of late-type
  stars.\\ 
\indent 
To date, notwithstanding a number of valuable projects, such as MONITOR (Hodgkin et al. 2006), EXPLORE/OC (von Braun et al. 2005),  and the  RCT monitoring project (Guinan et al. 2003), which are rapidly increasing our knowledge on the rotational
 properties of late-type members of open clusters,  the number of studied open clusters as well as the  number of variables in each cluster  have not been large enough to fully constrain the various models proposed to describe the mechanisms which drive the angular
 momentum evolution. Specifically, the sequence of ages at which the angular momentum evolution has been studied still has significant  gaps and the sample of cluster members for a number of clusters is not as complete as needed.\\
\indent
 The major science drivers  of the RACE-OC project are to explore rotational properties  of late-type stars in selected open clusters and associated stellar  activity. So far, there have not been many studies exploring the age  dependences of various manifestations of  stellar activity such as spot temperature, total spot area, the spatial  distribution of starspots on stellar surfaces, starspot activity  cycles, flip-flop phenomena and surface differential rotation (SDR). Therefore, in comparison to  similar ongoing projects, we focused our  attention  on the time evolution of stellar magnetic activity.\\
    \indent
    Keeping these objectives in mind, we have selected  open clusters with an age in the range from 1 to 500 Myr (Messina 2007a) for which no rotation and activity  investigations have been so far carried out. Furthermore, top priority is given to the open clusters which  fill the gaps in the empirical description of the age-activity-rotation relationship. We have also included in our sample clusters which have been already extensively studied, such as the Pleiades and the Orion Nebula clusters (Padmakar Parihar et al. 2008). The motivation behind this is  to further enrich the sample of periodic variables   and  to explore the long-term magnetic activity, e.g. to search for activity cycles and SDR, by making repeated observations of same clusters over several years.\\
 \indent
 Our sample also includes  open clusters  which were previously monitored  with different scientific motivations, but the re-analysis of archived  time series data  can provide very valuable results on late-type stars. This is the case of the  $\sim$500-Myr intermediate-age open cluster NGC\,2099 (M37), which is the first cluster of the RACE-OC project for which we present our results. Although it was previously studied to search for early-type pulsating variables (Kang et al. 2007), the collected observations allowed us to carry out a valuable and accurate study of rotation and magnetic activity. To date, there is no available data on rotation periods of clusters between the age range from  200 to 600 Myr and, hence, the study of NGC\,2099 with an age of  $\sim$ 500 Myr is one attempt to fill the gap of ages  while  modelling the angular momentum evolution (Barnes 2003; 2007).
\begin{figure}
\begin{minipage}{10cm}
\centering
\includegraphics[scale = 0.4, trim = 0 50 0 0, clip]{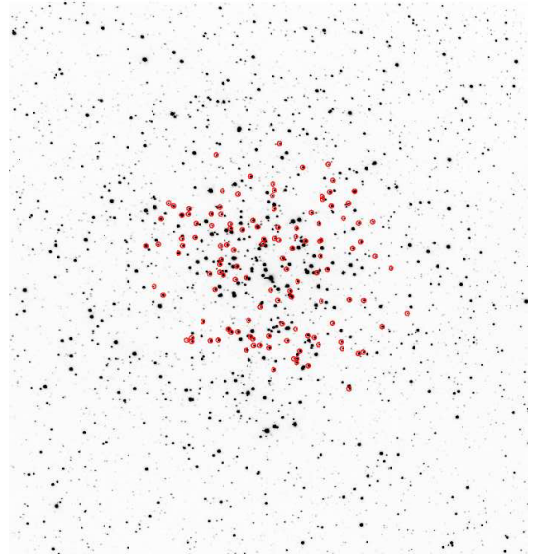}
\end{minipage}
\caption{\label{M37} Observed V-band CCD field (22.2 $\times$ 22.2 arcmin$^2$) of the open cluster NGC\,2099. Small circles identify the cluster candidate members newly discovered to be periodic variables.}
\end{figure}

NGC\,2099 (M37) is a $\sim$500 Myr intermediate-age  open cluster at distance of nearly 1400 parsec. The cluster, containing  thousands of members within core radius of 4-6 arc-minutes, is found to be very suitable for any monitoring program, using  intermediate aperture telescope  with moderate FoV. The cluster is very close to galactic plane ($l \sim 3^o$) and hence subjected to substantial reddening E(B$-$V) $\sim$ 0.2-0.3 mag. In the recent past, NGC\,2099 has been extensively studied by several researchers to obtain global parameters and to characterized its members (Mermilliod et al. 1996; Nilakshi \& Sagar 2002;  Hartman et al. 2007a and references therein).  With different motivations, NGC\,2099 was also chosen for photometric monitoring by three different groups  to search for new variables. The first two variability surveys carried out by Kiss et al. (2001) and Kang et al. (2007) could identify only 24 new variables. Whereas, the much deeper and very rigorous  MMT transit survey, in which   several hundreds of variables are identified, has been carried out  very recently by Hartman et al. (2007b). We started our  project on this cluster earlier than the MMT survey and before the Hartman et al. (2007b)  detailed work was communicated. Our work on NGC\,2099 can be considered valuable due to the following reasons, at least:
\begin{enumerate}
\item Stars brighter than 14.5 mag in r band ($\sim$ 15 mag in V band), that is F- and earlier-type stars, are saturated in MMT survey and hence our  bright
 NGC\,2099 F-type variables  are the first to be discovered and they help to make the sample of variable stars more complete for any future study.
\item  Observing epochs are different (2004 for present study  and 2006 for MMT survey) and hence the common  variables of both surveys can be utilized  to explore variations in light curves, related to magnetic activity.
\item  The prime motivation of present study is not just to identify the new variables in the cluster, but to use them to  investigate evolution of angular momentum and stellar activity vs.  age.
\end{enumerate}
  In Sect.\,2 we give information on observations, achieved photometric accuracy and selection criteria for membership. The search for periodic variables is presented in Sect.\,3 and results are given in Sect.\,4. Discussion and conclusions are given in Sect.\,5 and 6.

\section{Observations and data analysis}
The present study is based on observations taken in January 2004 with the 1.0m telescope at the Mt.\,Lemmon Optical 
Astronomy Observatory (LOAO) in Arizona (USA), which feeds a 2K$\times$2K CCD camera. The observed FoV is about 22.2 
$\times$ 22.2 arcmin$^2$ at the f/7.5 Cassegrain focus. A sequence of 581 images were collected in the V-band filter 
with an exposure time of 600 s over a total time interval of about 30 days. A detailed description of observations and 
data reduction was already given in the paper by Kang et al. (2007) to which we refer the reader. An example image of 
the observed NGC\,2099 field is plotted in Fig.\,\ref{M37}, where the over-plotted small circles identify the candidate 
cluster members  discovered by us to be  periodic variables.\\
\begin{figure}
\begin{minipage}{10cm}
\psfig{file=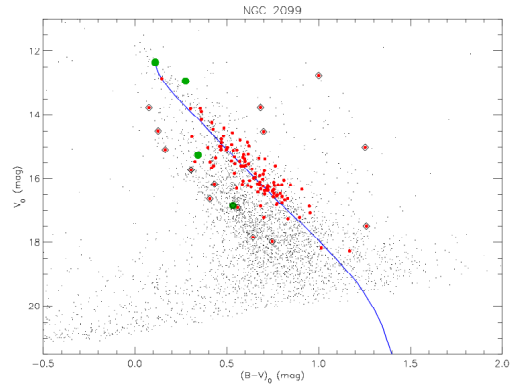,width=10cm,angle=270}
\end{minipage}
\caption{\label{HR} HR diagram of the stars (dots) detected in the field of NGC\,2099. Periodic variables are plotted with red small bullets. A diamond is over-plotted to periodic non-cluster members. Large green bullets represent  newly discovered candidate detached eclipsing binaries. The solid line represents the isochrone corresponding to an age of  t=485$\pm$28 Myr, E(B$-$V) = 0.227 mag and[M/H]  = 0.045 (Hartman et al. 2007a).}
\end{figure}
\begin{figure}
\begin{minipage}{10cm}
\centerline{
\psfig{file=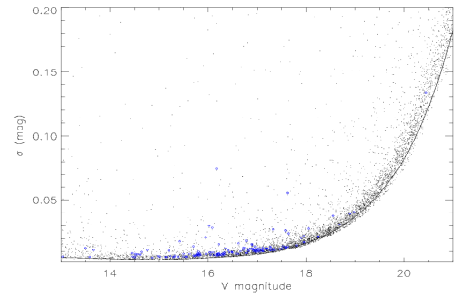,width=8cm,height=5cm,angle=270}
}
\end{minipage}
\vspace{1cm}
\caption{\label{accuracy} Standard deviation $\sigma$ of the light curves of the stars detected in the field of NGC\,2099 vs. average magnitude. The solid line represents a composite fit to the lower envelope of the  $\sigma$  distribution and gives the best achieved 
photometric accuracy. Small blue bullets represent the periodic variable candidate cluster members. }
\end{figure}
\indent
 The HR diagram of all the stars detected in  600-s exposures of the NGC\,2099 field is displayed as dots in
Fig.\,\ref{HR}. In the HR diagram  the newly discovered periodic variable stars are marked by small red bullets. Bullets 
with an over-plotted  diamond are  non-cluster member periodic variables, according to the selection criterion   of  
membership  outlined below. The large green bullets show the  detached eclipsing binaries newly discovered in our 
analysis. The solid line is the cluster isochrone. For the present analysis we adopted the following cluster parameters 
from Hartman et al. (2007a),  age = 485$\pm$28 Myr, E(B$-$V) = 0.227 mag, [M/H] = 0.045 and distance modulus (m$-$M) = 11.572 
mag. The proximity of any star in the HR diagram with respect to the cluster isochrone has been  adopted as criterion  
to select cluster members (see e.g. Irwin 2006, 2007). Any star with given intrinsic (B$-$V)$_o$ color and magnitude difference $\Delta$V smaller 
than $\pm 1.4$ mag with respect to the isochronal magnitude was selected as candidate cluster member. 
This magnitude difference is large enough to include the sequence of binary stars, if any, and to properly take into 
account the error on de-reddened (B$-$V)$_o$ color.
Such a  membership selection, based on photometry alone, cannot prevent from a certain level of contamination by 
non-cluster field stars. Hartman et al. (2007a) report such contamination to be on average about 39\%. However, as 
shown in Fig.\,\ref{M37}, the periodic variables we discovered are all in the  innermost NGC\,2099  field region, 
where the population of cluster members is usually relatively higher and, hence, the percentage contamination is expected 
to be significantly smaller than on average.\\
\indent
The 600-s long exposures of the NGC\,2099 field have allowed us to achieve a photometric accuracy $\sigma$ in the V band as good 
as 0.003 mag in the $13 < $V$ < 16$ magnitude range, and better than 0.01 mag for all stars brighter than V$\simeq$17.5 mag. 
In Fig.\,\ref{accuracy} we plot the  standard deviation of the light curves of all stars detected in the $13 < $V$ < 21$ magnitude 
range vs. their mean V magnitude. The lower envelope of the $\sigma$ distribution is populated by either non variable 
and  variable stars with least intrinsic variation. The lower envelope gives a measure of the best photometric accuracy 
we achieved at different magnitudes. 
Here we would like to point out that the $\sigma$ values are not those automatically computed by DAOPHOT in the PSF magnitude extraction, but time series data of each star were   binned with the time interval of 20 minutes. 
For each bin the mean magnitude and standard deviation were computed and, finally, averaged to obtain $\sigma$. The standard deviation computed following this procedure is 
 an empirical estimate of effective precision of our photometry. Such an estimate is conservative, because  the true 
observational  accuracy could be in principle even better for stars showing substantial variability  within the  time 
scale  closer to our fixed binning time interval.
  
 To construct the relation between observational accuracy and magnitude, following the procedure given e.g. by Roze \& Hintz 
(2007), the lower envelop of  Fig.\,\ref{accuracy}  was fitted  by a piecewise continuous function of the form:

\begin{figure*}
\begin{minipage}{18cm}
\centerline{
\psfig{file=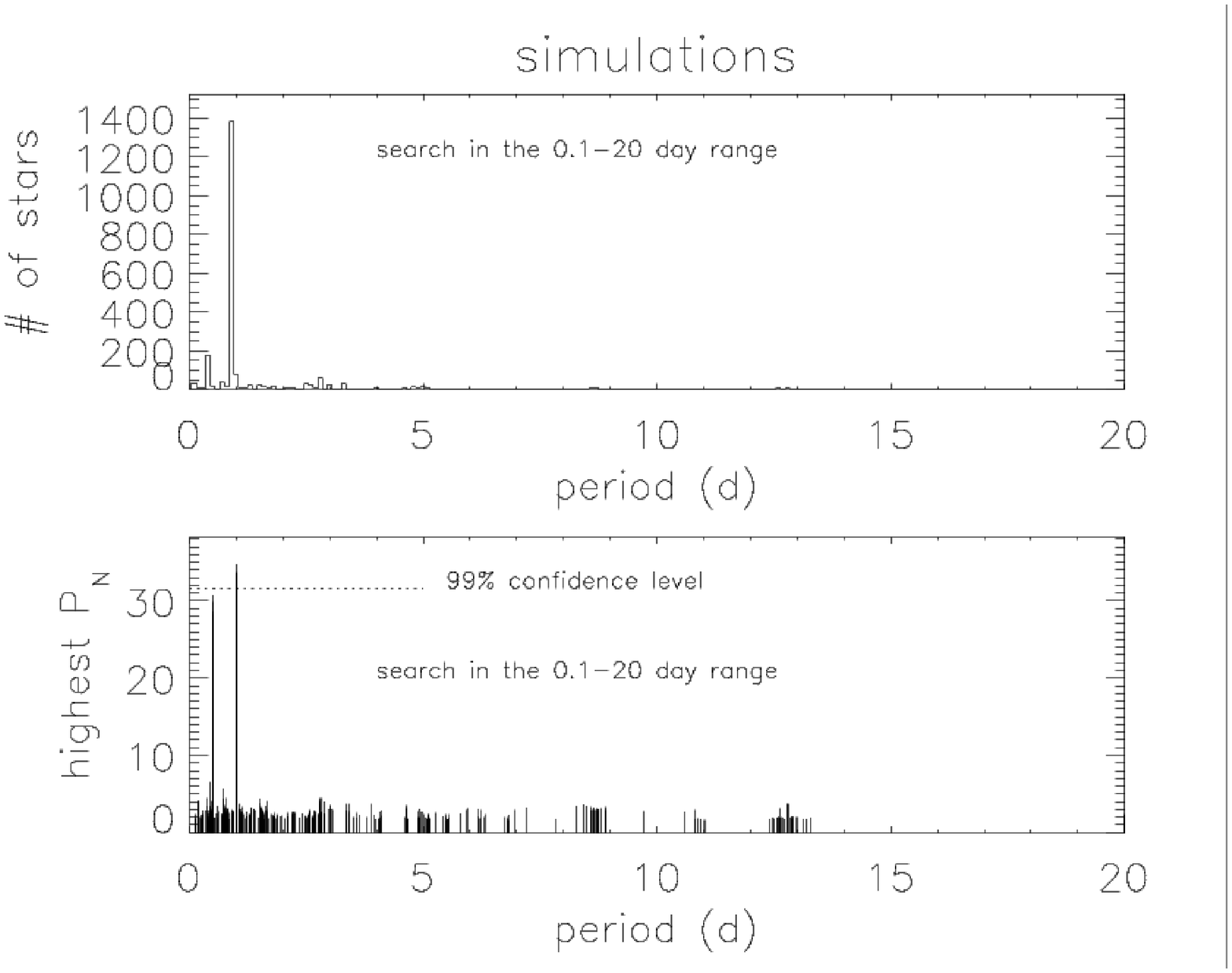,width=10cm,height=10cm,angle=270}
\psfig{file=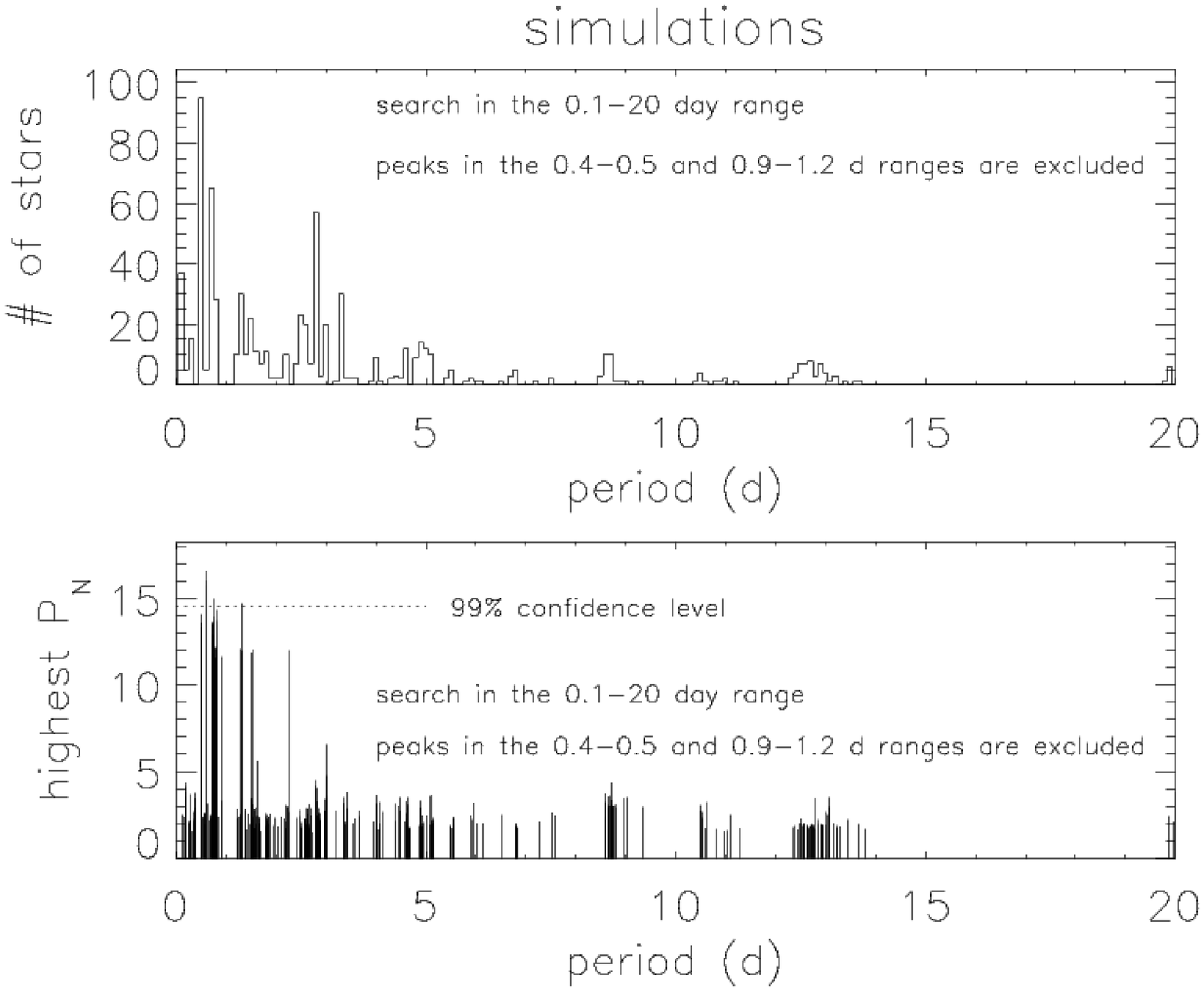,width=10cm,height=10cm,angle=270}
}
\end{minipage}
\caption{\label{distri_permu} \it Left panels: \rm results of periodogram analysis on ''randomized'' light curves.  \it Right panels: \rm the same as in left panel but with exclusion of power peaks in the 0.4-0.5 and 0.9-12 day ranges. }
\end{figure*}

\begin{tabular}{llr}
&\\
               & = $0.1495 \times e^{-0.26V}$   &   13 $<$ V $\le 15$\\
accuracy (mag) & = $3.5 \times  10^{-6} e^{-0.45V}$  &   15 $<$ V $\le  17$\\
               & = $9.2 \times 10^{-9} e^{-0.80V}$ &    V $>  17$\\

&\\
\end{tabular}
 Out of a total of 2250 stars detected in the $13<$V$<20$ magnitude range,  1746 turned out to be candidate cluster members.

\section{Search for periodicities}
One of the primary goals  of the present study is to search for periodicities in the time series photometric  magnitudes of all the members of NGC\,2099 open cluster. We have limited our investigation to stars in the $13<$V$<20$ magnitude range. Stars brighter than V$\simeq$13 are saturated in most of long-exposure frames, whereas,  stars fainter than V$\simeq$20 have a photometric accuracy rather low (lower than 0.07 mag) to allow us to reliably detect any periodicity, as it will better discussed in Sect.\,4.3. 
We expect that the presence of periodicity in the cluster members of G or later spectral type likely arises from solar-type magnetic activity.
Specifically, uneven distributions of cool spots along the stellar longitude, whose visibility is modulated by the star's rotation, give rise to quasi-periodic variation of the observed stellar flux. For these stars, the periodicity indeed represents the star's rotation period. For the candidate cluster  members of F spectral type, which are not expected to host magnetic activity, the presence of 
periodicity may likely arise from pulsations. In these stars, the periodicity indeed represents the period of one of pulsational modes.\\
\indent
Two different periodogram approaches, the Scargle-Press and the 
CLEAN periodograms, have been used to search for significant periodicities. In the following sub-sections we will briefly describe the techniques and  the criteria used to classify variable stars as periodic.

\subsection{Scargle-Press periodogram}
The Scargle technique has been developed in order to search for significant periodicities in unevenly sampled data (Scargle 1982; Horne \& Baliunas 1986). The algorithm calculates the normalised power P$_{\rm N}$($\omega$) for a given interval of angular frequencies $\omega = 2\pi\nu$. The highest peak in the calculated periodogram power spectrum reveals  the presence and the frequency of a
periodicity in the analysed  time series data.  In order to determine the significance level of the periodic signal, the height of the corresponding power peak is compared with  the false alarm probability (FAP). \\
\indent
The FAP is the probability that a peak of given height
is due to simply statistical variations, i.e.   white  noise. This method assumes that each magnitude measurement is independent from the other. However, this is not strictly true for our data time series (see, e.g. Herbst \& Wittenmyer 1996; Stassun et al. 1999) consisting of numerous data consecutively collected within the same night and with a time sampling much shorter than  both the periodic or the irregular intrinsic variability timescales we are looking for (P$^d$=0.1-20).  In order to overcome this problem,  we decided  to determine the FAP differently than proposed by Scargle (1982) and Horne \& Baliunas (1986), which is only based on the number of independent frequencies, but following the approach described by Herbst et al. (2002) based on Monte Carlo simulations. \\
\indent
Specifically, we selected a total of 2250 stars whose magnitude is in the 13 $<$ V $<$ 20 magnitude range and which were present in all 581 long-exposure images, thus securing that the time sampling is equal for all the stars under analysis.
The magnitude measurements and decimal parts of the MJD's of each star's time series were left untouched while the day numbers were randomly scrambled, thus securing that the correlations in the original data is preserved. Then, we applied the periodogram analysis to the "randomized'' data series of each star.
For each computed periodogram we retained the highest power peak and the corresponding period. In the top-left panel of Fig.\,\ref{distri_permu} we plot the distribution of detected periods from our simulations, whereas in the bottom-left panel we plot the distribution of the highest power peaks vs. period. \\
\indent
The FAP related to a given power P$_{\rm N}$  is set to be the fraction of randomized light curves which have the highest  power peak exceeding P$_{\rm N}$, which is the probability that a peak of this height is due simply statistical variations, i.e. white noise.  We find that the power corresponding to a FAP $<$ 0.01 is P$_{\rm N}>$ 31.64. However, we note that such a power threshold is period dependent and, in our specific case, is set by those peaks  appearing in the 0.4-0.5 and 0.9-1.2 day ranges, which are about 90\% of total detections from simulations.  
Therefore, P$_{\rm N}=$ 31.64 more exactly gives the 99\% level of confidence of a period detected in these mentioned time ranges. A straightforwardly application of this power level largely underestimates the confidence level of any other period detected outside these ranges with the risk to classify real detections as they were spurious.
Therefore, after removing all peaks falling in these mentioned ranges, we recomputed the distribution of power peaks vs. period. The results, which are plotted in the right panels of Fig.\,\ref{distri_permu}, show that a 99\% confidence level is reached for power levels  P$_{\rm N}>$ 14.60. In the following analysis these two different thresholds will be applied to the corresponding period ranges to select the final sample of periodic variables.

\begin{figure*}
\begin{minipage}{18cm}
\centering
\includegraphics[scale = 0.8, trim = 00 100 0 0, clip, angle=270]{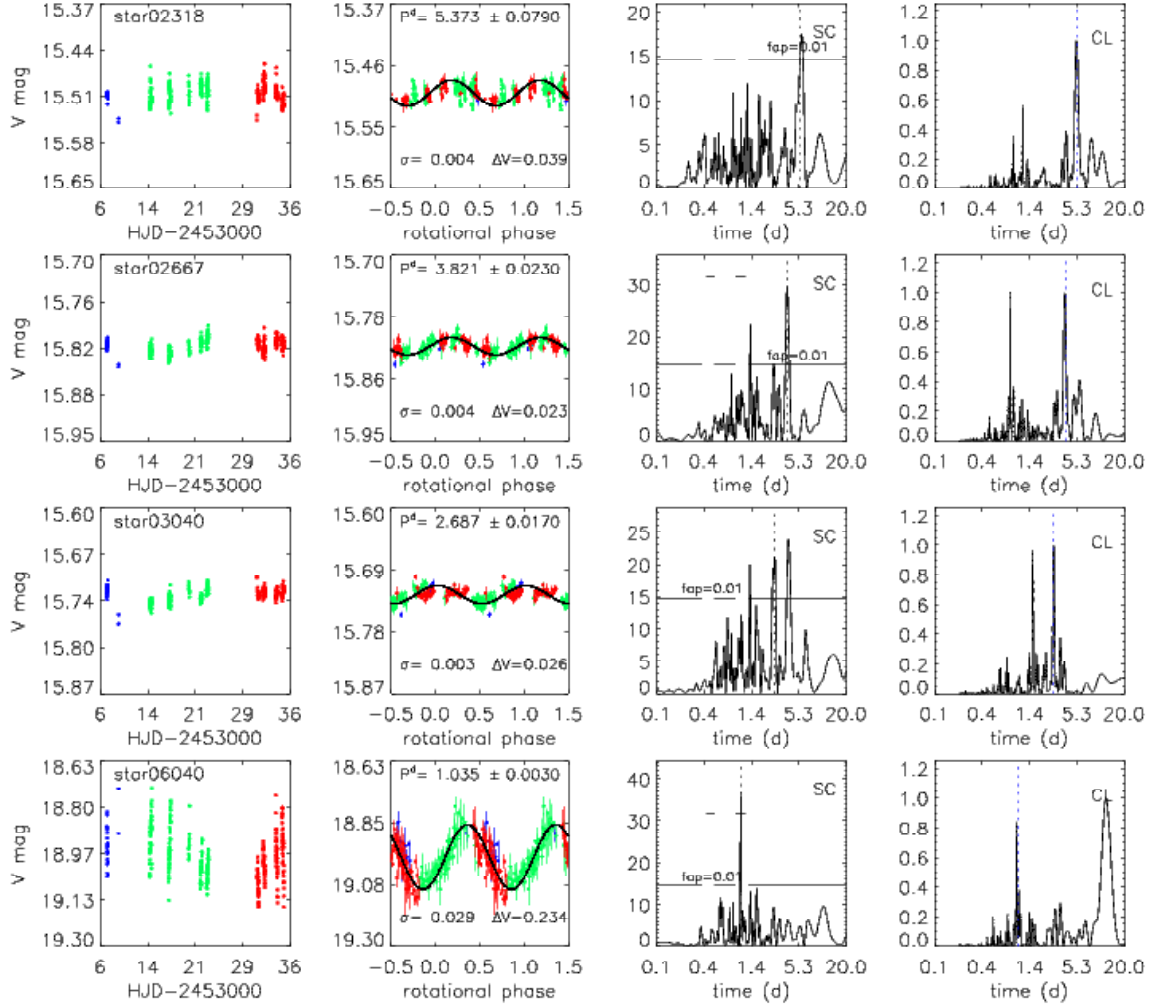}
\end{minipage}
\caption{\label{curve_example} Example of different results obtained with our period search analysis. Panels show (from left to right)  the data time series with color showing different time intervals; the phased light curve with labelled the rotation period, its uncertainty, the data accuracy and light curve amplitude. \bf Star 02318: \rm  only one period is detected 
with confidence level larger than 99\% and, independently confirmed  by CLEAN  analysis. 
\bf Star 02667:\rm two periods are detected; however, the shorter one is disregarded since it does not reach the power level of P$_{\rm N}=31.64$
set for the 0.9-1.2 day range. \bf Star 03040: \rm two periods are detected with confidence level larger than 99\%, however only the highest is confirmed by
the CLEAN analysis. \bf Star 06040:\rm only the  period detected is in the 0.9-1.2 day range and has a power level larger that the one set for this period range.}
\end{figure*}

\subsection{CLEAN periodogram}

The Scargle periodogram technique makes no attempt to account for the ob­servational window function W($\nu$), i.e. some of the peaks in the Scargle periodogram are result of the data sampling. This effect is called aliasing and even the highest peak could be an artifact. The CLEAN periodogram technique by Roberts et al. (1987) tries to overcome this shortcoming of the Scargle periodogram. For a detailed description we refer the reader to Roberts et al. (1987) or Bailer-Jones \& Mundt (2001).

\section{Results}
For all the candidate cluster member stars which turned out from our analysis to be periodic variable we list the target ID, RA and DEC coordinates (J2000.0), extinction-corrected V magnitude V$_o$ and color (B$-$V)$_o$ in the on-line Table\,1.
The results of our Fourier analysis are given in the on-line Table\,2, where we list the star's ID, the period detected by Scargle (P$_{\rm Scargle}$) and by CLEAN (P$_{\rm CLEAN}$) algorithms, the normalized peak power (P$_{\rm N}$), the adopted rotation period (P), i.e. the one among P$_{\rm Scargle}$ and P$_{\rm CLEAN}$ giving the less scattered phased light curve, and its uncertainty as computed following Horne \& Baliunas (1986), the average V magnitude ($<$V$>$), the light curve standard deviation ($\sigma_{\rm TOT}$), the achieved photometric accuracy ($\sigma$), the light curve amplitude ($\Delta$V). The amplitude of the light curve was computed by making the difference between the median values computed by considering the upper and lower 15\%  of magnitudes of the light curve (see, e.g. Herbst et al. 2002). That allows us to prevent from overestimating the amplitude due to possible outliers. We  also list  the total number of observations, a note about either the membership or its ID for already known variables.\\
\indent
In Fig.\,\ref{curve_example} we display a selection of the most common  results obtained with our period search analysis. In the panels from left to right we plot the  data time series (V-band data vs. MJD).  Different colors are used to distinguish three different datasets collected at different time intervals of the observing run. This helps to disentangle the scatter in the data arising from three different sets of observing runs. In fact, we noticed that a number of stars undergo changes in the light curve shape and amplitude on a time scale shorter than the whole time interval of observations (about 30 days). 
In the second panel from left we plot the light curve phased with the period reported in the label along with its uncertainty. We also report the data accuracy ($\sigma$) and the light curve amplitude ($\Delta$V).
 In the third panel we plot the Scargle periodogram where the horizontal solid line represents the power level corresponding to the 99\% confidence level. We note that in the 0.4-0.5 and 0.9-1.2 day range such level is  to a much higher power (P$_{\rm N}$=31.64). The dashed vertical line indicates the selected periodicity. 
The right-most panel displays the power spectrum from CLEAN analysis. \\
\indent
We have cases, such as for star 02318, in which only one period has a confidence level larger than 99\%, which  is independently confirmed by the  CLEAN analysis. There are cases, such as for star 02667, where two periods are detected. However, the shorter one is disregarded since it does not reach the power level of P$_{\rm N}=31.64$
which we set for the 0.9-1.2 day range. There are other cases,  such as for star 03040, where two periods are detected with confidence level larger than 99\%. However,  only one is confirmed by the CLEAN analysis. Finally, there are cases, such as for star 06040, where the  period is detected in the 0.9-1.2 day range and has a power level larger that the one set for this period range.
For 10 stars two different periodicities of comparable power were detected by both Scargle and CLEAN periodograms. In these cases we considered to be more reliable  the period with higher power and which produces the less scattered phased light curve and marked the second period to be an alias. In most case the second periodicity represents a beat period (B) which is related to the rotation period (P) by the following relation:
$$ \frac{1}{B}=\frac{1}{P}\pm 1$$
It can be  identified to be an alias of the rotation period, since it produces a larger scatter in the phased light curve. 
We could detect periodicities with high confidence level ($>$99\%) in a total of 135 stars. \\
\indent 
We found 135 periodic variables, 16 of which were previously discovered by Kiss et al. (2001) and Kang et al. (2007). The total number of periodic cluster members newly discovered by us is  106, which is the final sample analysed in the following sections. \\
\indent
In the on-line Figs.\,11-13 we plot magnitudes time series,  phased light curves and power spectra of the known variable V3-V7 discovered by Kiss et al. (2001) and KV3-KV17 discovered by Kang et al. (2007). We detected no evidence of eclipses for the variables V1 and V2, confirming the results of Kang et al. (2007), whereas the variables KV1 and KV2 got saturated in the long-exposures analysed by us. For 14 out of 20 remaining known periodic variables our present analysis could confirm the previously determined periods. For 6 out of 8 $\delta$ Scuti stars we could detect the primary frequency. Six of of 20 known variable stars did not pass our period detection criteria, i.e. periodicity was detected with low confidence level. For all the known contact eclipsing binaries our analysis detected the highest peak with greater than 99\% confidence level at exactly half the orbital period. In fact the Fourier analysis done by us is better suited for single-peaked light variations which are fitted by a single sinusoid function. In the present case, since the known contact eclipsing binaries have two minima of comparable depth, they have been all fitted by a single peaked variation of half orbital period. The very good agreement between the periods detected by us and the previously known periods for the common periodic variables makes us confident on the reliability of the periodicity measured for all the stars in our sample.

\section{Discussion}
\subsection{F-type stars}
\begin{figure}
\begin{minipage}{10cm}
\centerline{
\psfig{file=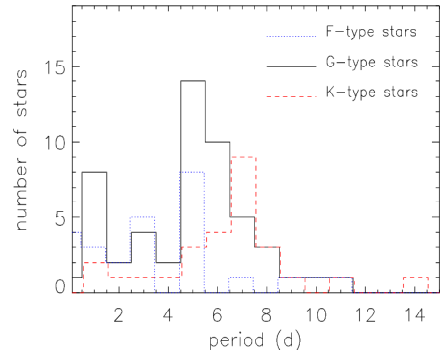,width=8cm,height=6cm,angle=270}
}
\end{minipage}
\vspace{1cm}
\caption{\label{histogram} Distributions of  periods of the NGC\,2099 candidate cluster members. }
\end{figure}

As mentioned in Sect.\,1, a search for variability among F-type stars in the  NGC\,2099 cluster was the prime motivation of the previous analysis carried out by Kang et al. (2007). That analysis allowed to discover a total of 24 periodic variable stars including the  seven variables  already identified by  Kiss et al. (2001). Out of  17  newly discovered variables by Kang et al. (2007), 9 variables are identified as $\delta$ Scuti-type pulsating stars, 7 are found to be contact eclipsing binaries, and one is a peculiar variable star. Although the present paper is focused on G- and K-type variable stars, we nonetheless applied our Fourier analysis techniques briefly described in Sect.\,3 also to the sub-sample of F-type candidate cluster members. Our new analysis has allowed us to discover 26 new periodic variables among candidate cluster members. Their light curves along with their Scargle and CLEAN periodograms, are shown in the on-line Fig.\,14-16. The list of periodic F-type stars is given in Table\,2 with indication of the variable stars already discovered in previous studies (Kiss et al. 2001; Kang et al. 2007). \\
\indent 
The periodic variation of star light in  F-type stars  most  likely arises from pulsations. About 70\% of F-type variables have been found with periods too long (P$^d$ $>2$) to be likely attributed to pulsational modes (Fig.\,\ref{histogram}). One possibility is that these long-period F-type variables may not be cluster members, but they are foreground field stars. Early G-type field stars may have been designated as cluster F-type variables due to reddening over-correction. Consistently with this hypothesis, the B$-$V color of these stars is indeed at the boundary between inactive late F-type and active early-G stars. It is possible that further investigation of these stars may reveal them to be active G-type field stars.
 
\begin{figure*}
\begin{minipage}{10cm}
\centering
\includegraphics[scale = 0.7, trim = 0 0 0 20, clip, angle=270]{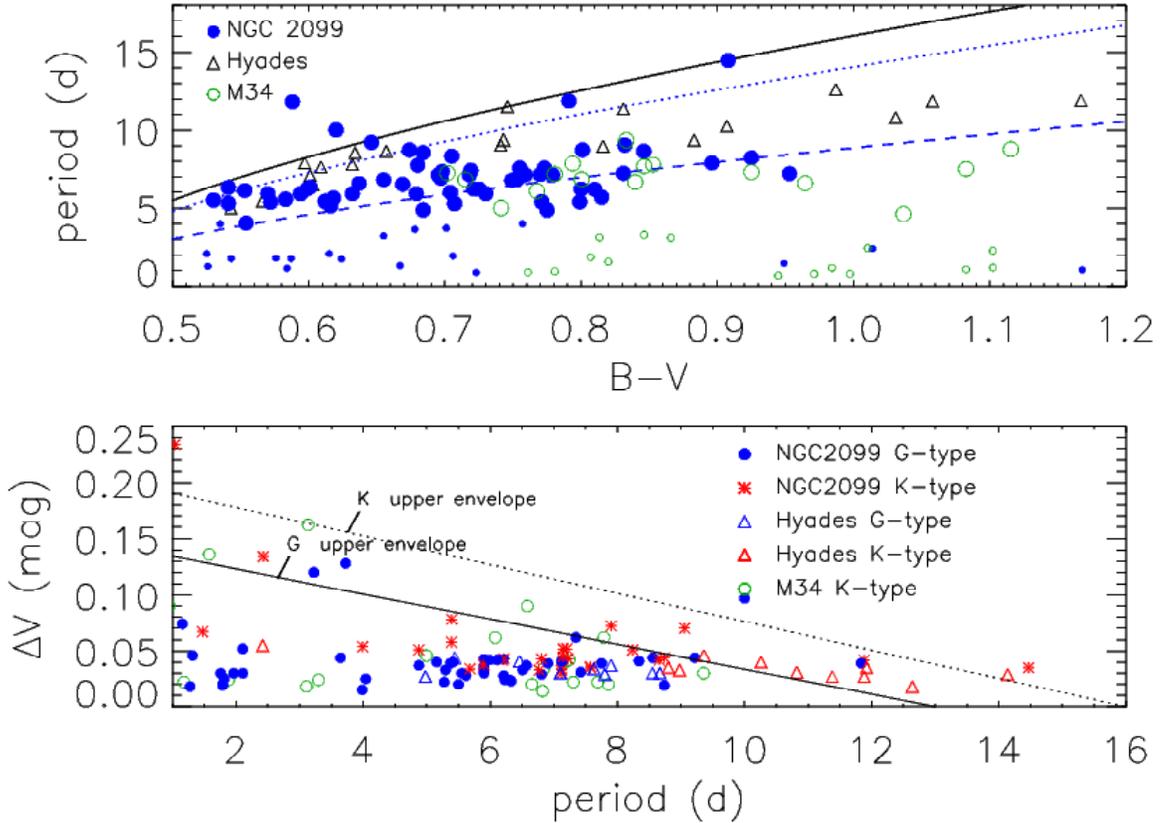}
\end{minipage}
\caption{\label{ampl} \it Top panel\rm: Rotation period vs. B$-$V color for NGC\,2099  (filled blue bullets), Hyades (open triangles), and M34 members (open green bullets) with overplotted the family of age-parameterized curves from gyrochronology corresponding to ages of 600, 500 and 200 Myr (from top to bottom, respectively). Fast rotators with P$^d$$<$ 4 among NGC\,2099 and M34 members are plotted with smaller symbols. \it Bottom panel\rm: light curve amplitude ($\Delta$V) vs. rotation period for NGC\,2099  G-type (blue bullets) and K-type (red asterisks) members, G-type and K-type Hyades members (open blue and red triangles, respectively), and M34 members (green open bullets). The upper limits of the $\Delta$V-P relation for dwarf  G (solid line) and K  stars (dotted line) from Messina et al. (2003) are over-plotted. }
\end{figure*}

\subsection{G- and K-type main-sequence stars}
The primary goal of  the present study is to investigate the rotational and magnetic activity properties of late-type (G-K)  members of NGC\,2099. The Fourier analysis has allowed us to newly discover the periodicity in 52 G- and 28 K-type stars, whose masses cover approximately the $ 0.7 \la M/M_{\odot} \la 1.1 $ range. The lower percentage of periodic K stars has to be mostly ascribed to the lower photometric accuracy achieved for K stars with respect to G stars. The results of our Fourier analysis are summarized in the on-line Table\,2, whereas the light curves, along with their Scargle and CLEAN periodograms, are plotted in the on-line Figs.\,17-26. \\
\indent 
For these late-type stars the detected periodicity most likely represents the stellar rotation period. It is believed that in late-type stars the observed variability arises from non-uniformly distributed cool spotted regions on the stellar photosphere, which are carried in and out of view by the star's rotation. In Fig.\,\ref{histogram} we plot the distribution of rotation periods for G- and K-type candidate cluster members. We note that our sample og 52 G stars does not incluse the 10 W UMa-type eclipsing binaries, whose rotation period is altered by tidal synchronization, and subjected to a different
  magnetic braking than in single stars. 
A Kolmogorov-Smirnov test reveals that G and K stars have different distributions, having K-type stars on average longer rotation periods than G-type stars. The distribution of G stars is peaked at a median period of  P=5.9 days, whereas the distribution of K stars at a median period of P=7.1 days.

In  Fig \,\ref{histogram} we  see that  a secondary peak exists of fast rotating stars (P$^d$$<$ 3), consisting of about 20\% of G stars. Such fast rotators are usually found in young open clusters but not expected  in a cluster with an age of about 500 Myr or older, which is the case of NGC\,2099 (Barnes 2003; 2007). There are several possibilities for the presence of these fast rotators.  Since the cluster is  relatively far away (1.4 kpc) and
 its galactic latitude is just 3 degrees, it may be contaminated by very young WTTS,  which are usually fast rotators. Alternatively, some of these fast rotators may be cluster members but belonging to close binaries of BY Dra, RS CVn, W UMa or FK Com types, which are rapidly rotating due to tidal interaction and synchronization. Third possibility may be the  false  detections of the periods. For example, the presence of two activity centers in the stellar photosphere at the epochs of observations, about 180$^{\circ}$ away in longitude from each other, can produce a light modulation with half of the true rotation period. In these cases, consecutive observation seasons are needed to detect the true rotation period (Padmakar Parihar et al. 2008). Identification of single fast rotating cluster members at an age of about 500 Myr would be indeed very interesting and challenging to models of evolution of stellar angular momentum.

\indent 
 Our study of NGC\,2099 is aimed at filling the gap in the description of the angular momentum and magnetic activity evolution in the age range from about 200 to 600 Myr. In the top panel of Fig.\,\ref{ampl} we plot the rotation periods  of the NGC\,2099 G and K stars against their reddening corrected B$-$V color. We also include  similar data  of the slightly older ($\sim$625 Myr; Perryman et al. 1998) Hyades open cluster from Radick et al. (1995) and of the younger ($\sim$200 Myr; Irwin et al. 2006) M34 cluster for a comparison. Such a comparison is restricted to the $ 0.7 \la M/M_{\odot} \la 1.1 $ mass range, which roughly corresponds to the 0.52 $<$ (B$-$V) $<$ 1.0 color range within which most of the NGC\,2099 candidate members fall. The (V$-$I) colors of M34 members were converted into B$-$V colors by interpolating between colors of dwarf standard stars as tabulated in Cox (2000).
Among the NGC\,2099 and M34 members, we can identify a longer-period branch (P$^d$$>4$), which is represented by filled and open bullets in the top panel of Fig\,\ref{ampl}, which is also present among Hyades stars (open triangles), and a short-period branch (P$^d$$<4$) which has no counterpart among Hyades stars. \\
\indent
Within the longer-period branch, we have computed the median period of G stars (0.52 $<$ B$-$V $\le$ 0.74) in NGC\,2099 and Hyades clusters, obtaining the following values: P$^d$$_{\rm Hyades}=7.6$ and P$^d$$_{\rm NGC2099}=5.4$. Unfortunately, only three stars in M34 fall within this color range to compute  a median value.
The same was done for early K stars (0.74 $<$ B$-$V $\le$ 1.0) in all three clusters, obtaining the following values: P$^d$$_{\rm Hyades}=10.3$, P$^d$$_{\rm NGC2099}=7.1$ and P$^d$$_{\rm M34}=7.2$. From these median values we infer that G stars have undergone a significant spin down in the age range from $\sim$500 to $\sim 600$ Myr, as expected from magnetic breaking. The unexpected result is that K stars have kept their rotation period  almost constant in the age range from $\sim$200 to $\sim 500$ Myr, then they have undergone a significant spin down only in the following 100 Myr, i.e. until the Hyades age. Unfortunately,  we do not have enough data for M34 to check whether  the rotation period of G stars has remained constant  over the age range from $\sim$200 to $\sim$500 Myr or not.
The difference in the rotation distribution of NGC\,2099 stars with respect to Hyades stars   is what is expected from the theory of angular momentum evolution, where the braking mechanism works more efficiently in K-type than in G-type stars. What is very different from theoretical expectations is that early K-type stars in NGC\,2099 show the same rotation distribution as in M34, that is they seem to have undergone no braking during the course of about $\sim$300 Myr. We have analysed our finding also in the framework of the gyrochronology proposed by Barnes (2003; 2007). In Fig.\,\ref{ampl} we plot the age-parameterized family of  theoretical curves corresponding to nominal ages of  the three clusters,  that is $\sim$ 625 Myr for Hyades (Perryman et al. 1998),  $\sim$ 500 Myr for NGC\,2099 (Hartman et al. 2007) and $\sim$ 200 Myr for M34 (Meynet et al. 1993), which are plotted as solid, dotted and dashed lines, respectively.
 Whereas G stars in NGC\,2099 seem to be in the 
 age range between $\sim$200 and $\sim$600 Myr, the early K stars are well fitted by the 200-Myr isochronal line, i.e. they appear to be \it rotationally unevolved \rm from the age of M34 ($\sim$200 Myr). \\
\indent
Our  measurements of rotation periods are very important since there is no determined  period distribution  at the age of about 500 Myr.  In the most recent work on this subject (Irwin et al. 2007) the distribution of rotation periods in the 200-600 Myr range was just interpolated. Now,  we have added  an additional point to better constraining evolution models. On the other hand our finding raises the question why magnetic braking, at least in G8-K5 stars, has been ineffective in the age range from about 200 to 500 Myr. Presently, we are not in the position to give any reasonable explanation; however, once the rotation period distributions for the other target clusters of RACE-OC will be determined, we will be able to carry out a more accurate and robust comparison between observation and theory. \\
\indent
In the bottom panel of Fig.\,\ref{ampl} we plot the light curve amplitudes vs. rotation period for the NGC\,2099, Hyades and M34 clusters. As investigated by Messina et al. (2001, 2003), the maximum  amplitude of light curve monotonically decreases with rotation period following power laws with different exponent in different mass ranges. Here, we over plot the fits from the cited works. We see that the  amplitudes of periodic variables of NGC\,2099 and M34 are always smaller than the maximum expected amplitude. One explanation of these low amplitudes is that it is very unlikely to find a star at its maximum light curve amplitude from only one observing season. One needs  10-20 light curves collected over years  to obtain  a maximum amplitude for spotted variables (see e.g. Messina et al. 2003; Messina 2007b). But, at the same time, we argue that  we  have 31 variables stars within, e.g., the P=5-7 days rotation interval and  over this period range, on a statistical basis  at least, 
 few stars are expected to show their  maximum light curve amplitude, which we never find. 
 Therefore, we  suspect that there should be some other age-dependent quantity which controls
the level of activity, other than  rotation and mass,  and makes older stars less active than younger. A similar suspect was already raised by Messina et al. (2001), who found evidence that, for a fixed mass and rotation period, the level of starspot activity increases from the zero-age main sequence up to the Pleiades age ($\sim$130 Myr) and then after it decreases  with age. At an age of  200 Myr, the amplitude of K stars is significantly decreased with respect to K-type Pleiades stars, as shown in the Fig.\,16 of the paper by Irwin et al. (2006). Again, the younger G8-K5 stars in M34 show a median light curve amplitude, that is a level of photospheric magnetic activity, which is larger than the older NGC\,2099 stars, although the mass range as well as the rotation period distribution are the same.

\begin{figure}
\begin{minipage}{10cm}
\psfig{file=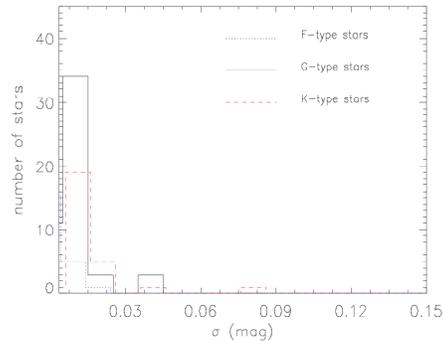,width=8cm,height=6cm,angle=270}
\end{minipage}
\vspace{1cm}
\caption{\label{sigma_per} Distributions of standard deviations of V-band light curves of  periodic variables found in the open cluster NGC\,2099. }
\end{figure}

\begin{figure}
\begin{minipage}{10cm}
\psfig{file=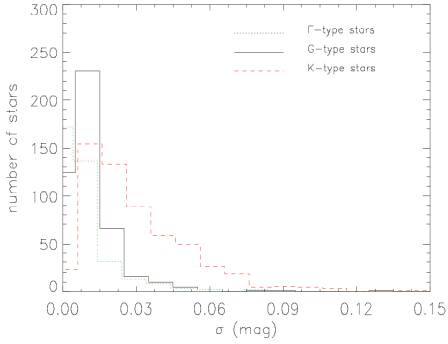,width=8cm,height=6cm,angle=270}
\end{minipage}
\vspace{1cm}
\caption{\label{sigma} Distributions of standard deviations of V-band light curves of stars identified as a non-periodic cluster variables. }
\end{figure}
\subsection{Non-periodic variables}
By analyzing the sample of G and K periodic stars, we find that the average ratio between the  V-band light curve amplitude and its standard deviation is 2.7$\pm$0.1, which must be considered to be valid for a type of variability  arising from cool starspots. The expected  maximum light curve amplitude  of G and K stars is $\Delta$V$\simeq$0.15 and $\Delta$V$\simeq$0.20 mag, respectively (see Fig.\,\ref{ampl}). Therefore, stars having an observational accuracy better than $\sigma\simeq$0.07 mag ($\simeq$0.20/2.7), which correspond to stars brighter than V$\simeq$20 (Fig.\,\ref{accuracy}), can be selected to search for  periodic variability.  Indeed, such accuracy allows us to detect with 3$\sigma$ confidence level, a variability of amplitude of $\simeq$0.20 mag, which is the maximum expected light curve amplitude for K-type stars.\\
\indent
Our  Fourier analysis could not determine any significant periodicity for  1746 cluster members  in the $13.0 < V < 20.0$ magnitude  range.  Nonetheless, some information on their nature of variability and membership can be derived from the analysis of their standard deviations.
In Fig.\,\ref{sigma_per} we plot the $\sigma$ distribution of  V-band light curves of  periodic  cluster variables. The median values  of the plotted  distributions are 0.013 and 0.016 mag
for G and K-type stars, respectively, which is consistent with
the behaviour shown by spotted stars. Although K stars on average rotate slower than G stars, they display larger photometric variability having deeper convection zones, more efficient dynamos, and consequently a larger activity level. In Fig.\,\ref{sigma} we plot the $\sigma$ distribution of all non-periodic  cluster variables. The median values  of $\sigma$ for G and K-type stars are found to  0.013 and 0.028 mag, respectively. The distributions of  $\sigma$ of periodic and non periodic G stars are similar, whereas for K-type stars the non periodic stars have larger  $\sigma$ and a tail toward larger values, but they never exceed the  0.07 mag limit expected for active spotted stars.
The discrepancy in the distribution between periodic and non-periodic K-type stars may arise from contamination of non-cluster stars, which we can guess are still spotted stars, but more active and, therefore younger and/or faster rotating. 

\subsection{Individual stars}
\begin{figure*}
\begin{minipage}{18cm}
\centerline{
\psfig{file=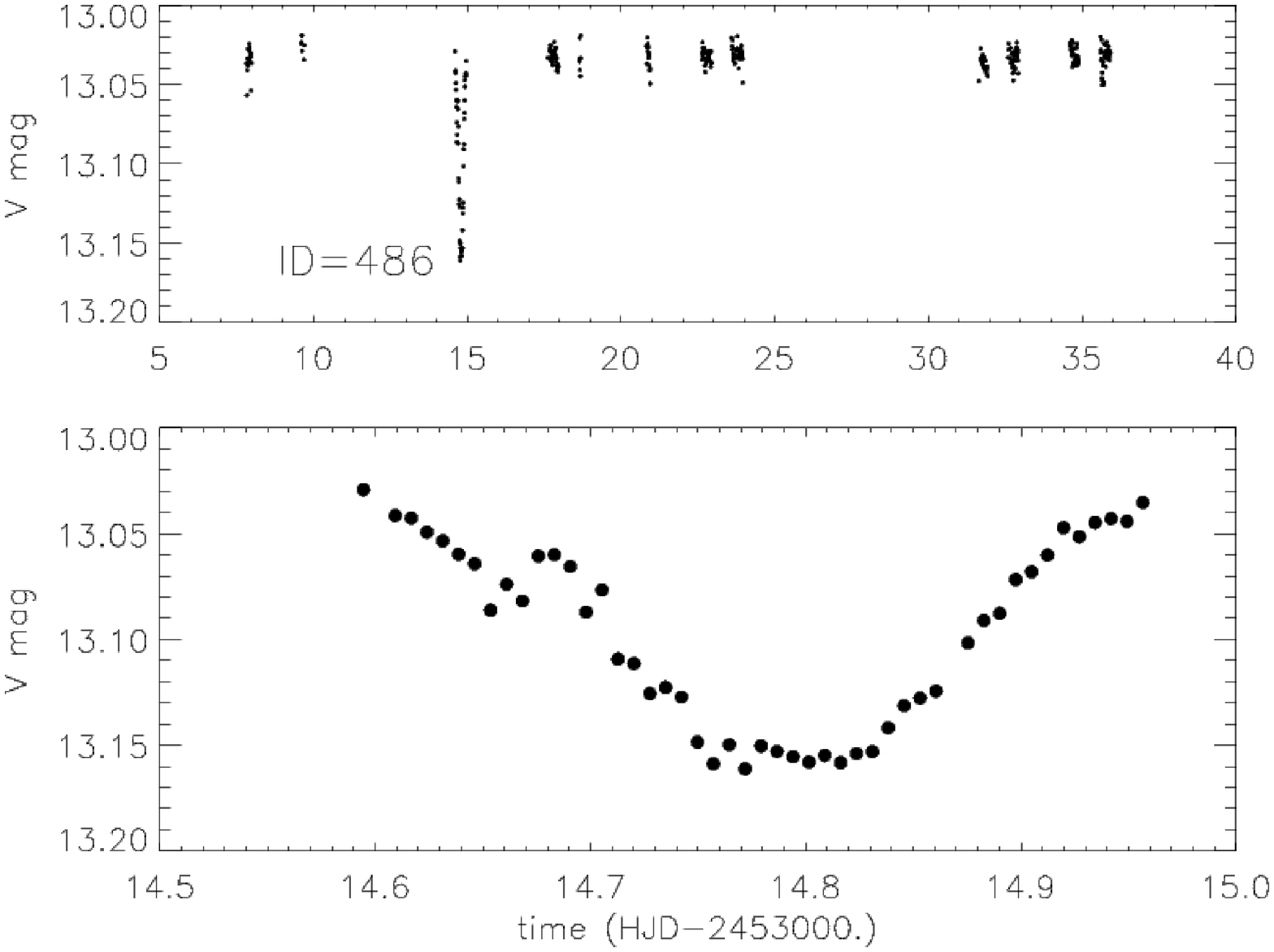,width=9cm,height=6cm,angle=270}
\psfig{file=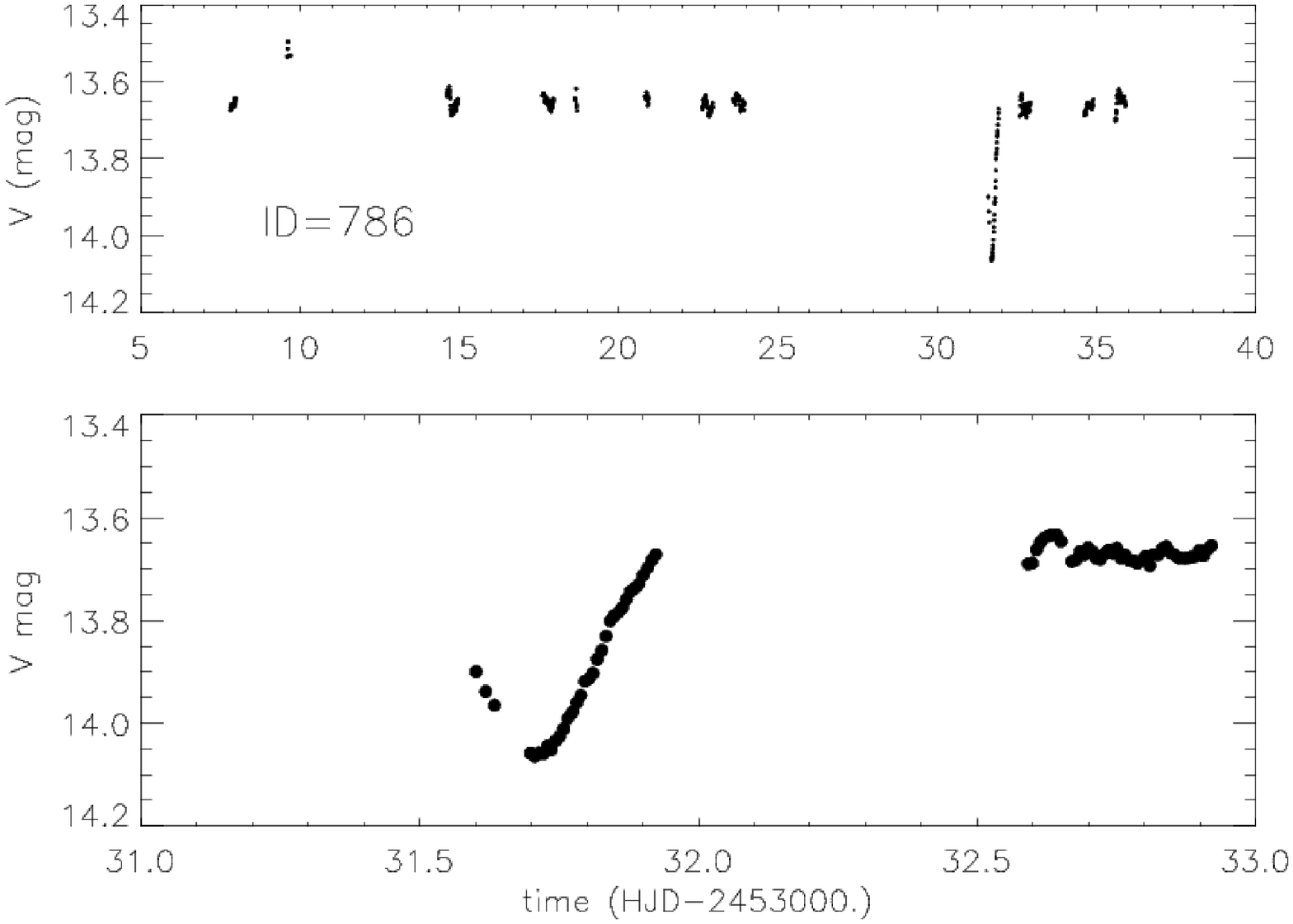,width=9cm,height=6cm,angle=270}
}
\end{minipage}
\begin{minipage}{18cm}
\centerline{
\psfig{file=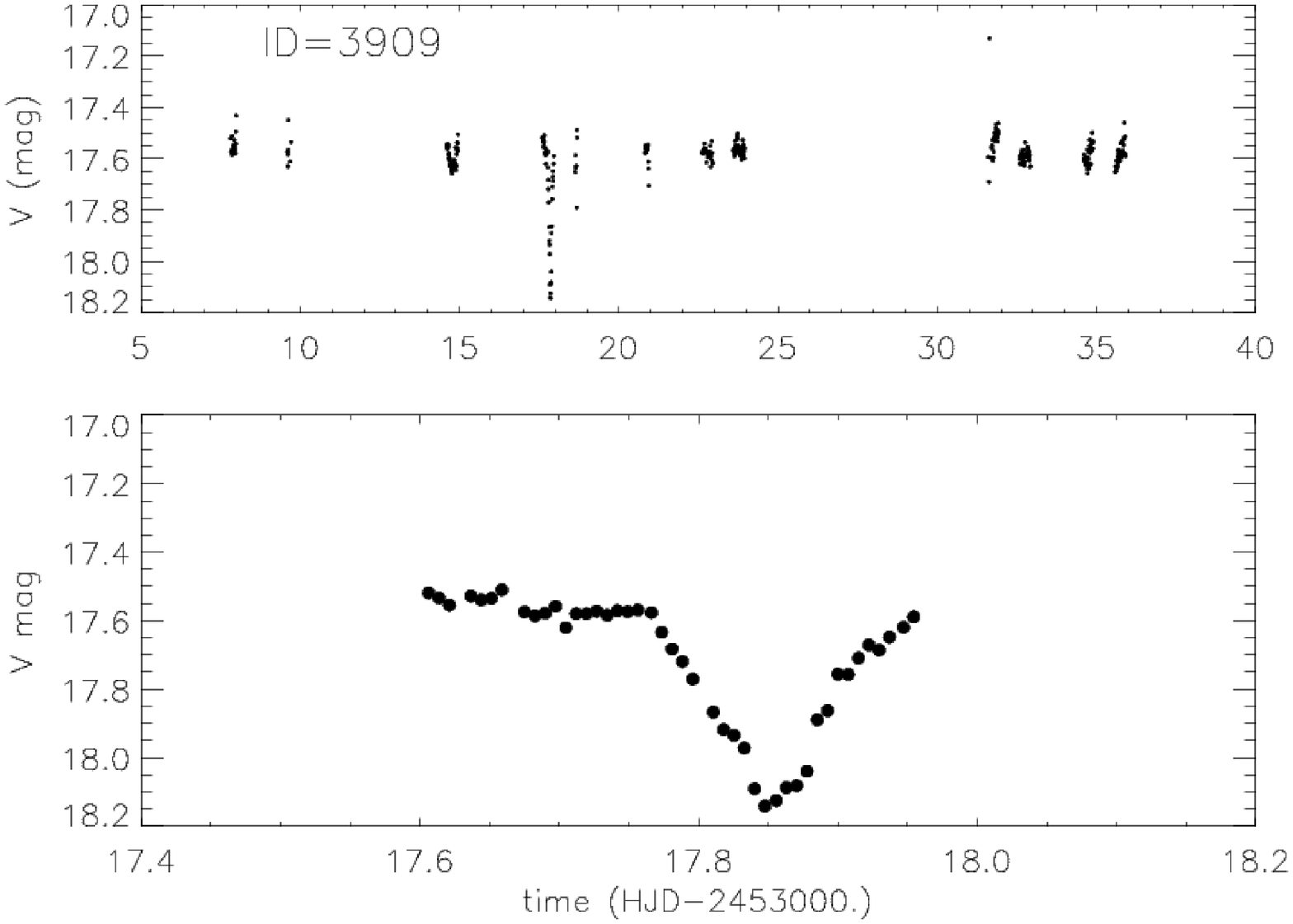,width=9cm,height=6cm,angle=270}
\psfig{file=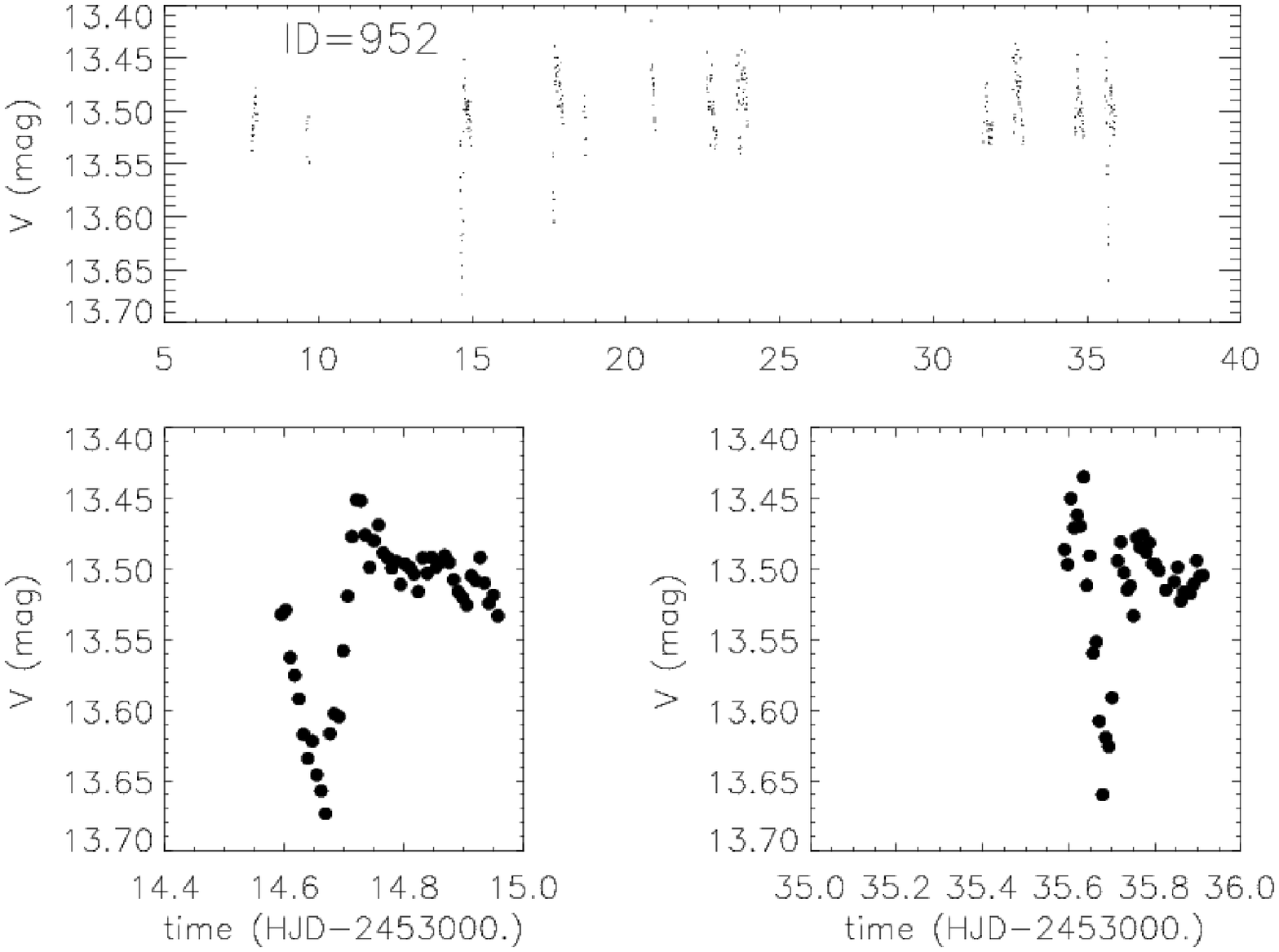,width=9cm,height=6cm,angle=270}
}
\end{minipage}
\begin{minipage}{18cm}
\centerline{
\psfig{file=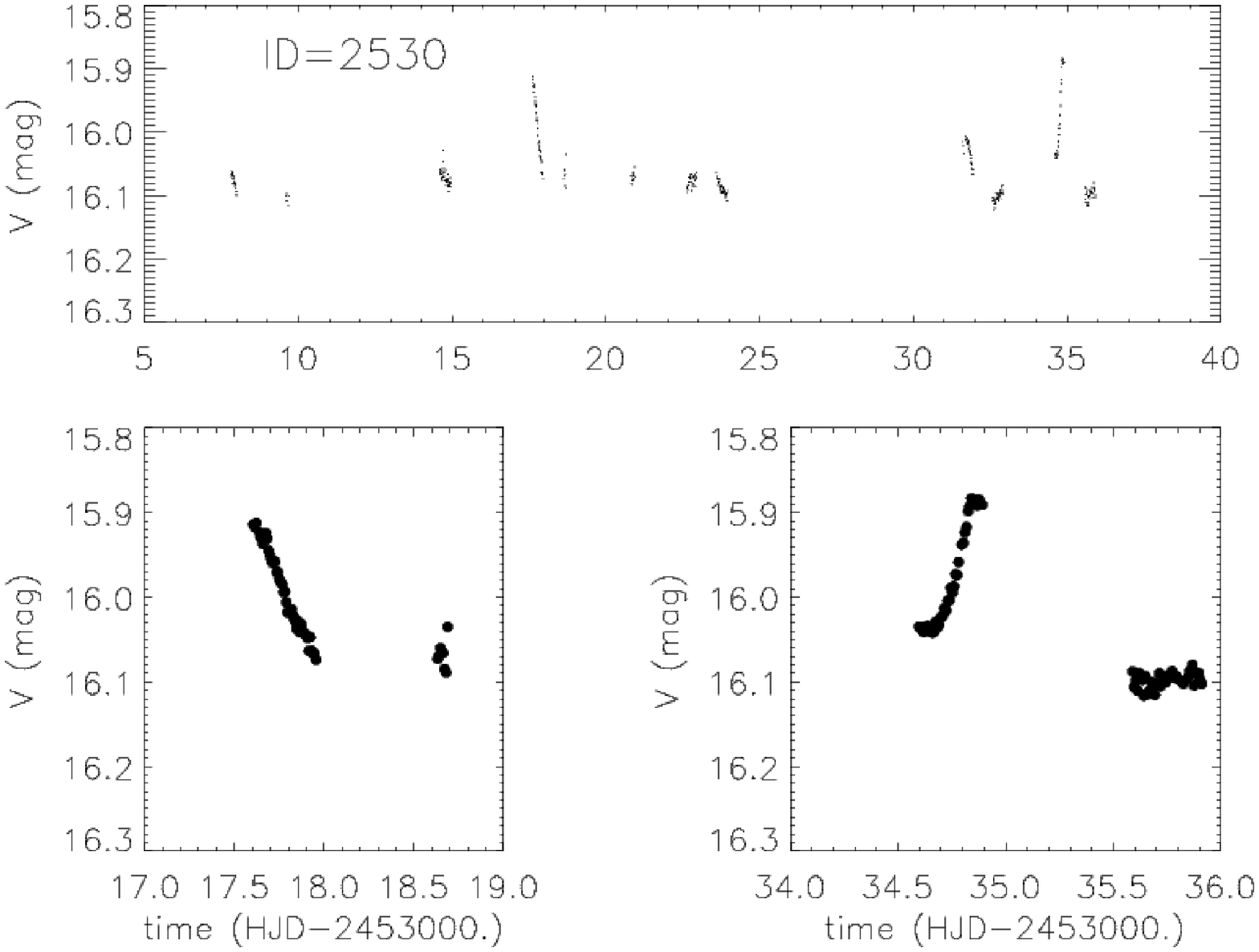,width=9cm,height=6cm,angle=270}
}
\end{minipage}
\caption{\label{eb} Complete V-band magnitude time series (upper panels) of five interesting stars among newly discovered cluster variables. Lower panels give an enlarged view of the most interesting nights.  Stars  ID=486, ID=786, ID=3909 and ID=952 are likely detached eclipsing binaries.  ID=2530 shows outburst-like brightening up to 0.15 mag.}
\end{figure*}

Our analysis has allowed us to discover 5 variables which certainly deserve further observations to characterize their nature and origin of variability. Four out of five variables are likely detached eclipsing binaries (Fig.\,\ref{eb}): ID=486, ID=786, ID=3909, and ID=952. For ID=486 (V$\simeq$12.4, B$-$V$\simeq$0.12) we could observe the full eclipse on HJD=2453014 of $\Delta$V $\simeq$ 0.13 mag, of about 8.9 h of duration and with the minimum occurring at HJD$_{\rm min}$=2453014.81. For ID=786 (V$\simeq$13, B$-$V$\simeq$0.30) we could observe one incomplete eclipse on HJD=2453031 of $\Delta$V $\simeq$ 0.53 mag, of about 12 h of duration and with the minimum occurring at HJD$_{\rm min}$=2453031.70. For ID=3909 (V$\simeq$16.9, B$-$V$\simeq$0.54) we could observe one almost complete eclipse on HJD=2453017 of $\Delta$V $\simeq$ 0.65 mag, of about 3.9 h of duration and with the minimum occurring at HJD$_{\rm min}$=2453017.86. For ID=952 (V$\simeq$12.8, B$-$V$\simeq$1.), although
the data are quite noisy, we could observe two almost complete eclipses on HJD=2453014 and 2453035 of similar $\Delta$V $\simeq$ 0.18 mag, of about 2.5 h of duration and with the minima occurring at HJD$_{\rm min}$=2453014.67 and 2453035.68. 
These  detached eclipsing binaries candidates are very important since three of them  are candidate cluster  members, according to our selection criterion. Therefore, they allow to make accurate determination of mass and radius and, very importantly, a more constrained comparison of various stellar  evolutionary models. Finally, ID=2530 (V$\simeq$15.4, B$-$V$\simeq$0.45) shows two outburst-like events in different nights HJD=2453017 and 2453034, which last several hours and during which it gets brighter up to 0.15 mag. The origin is quite intriguing because  the shape of the light curve at the brightening phase, as well the star's spectral type are not consistent with a flare-like event.

\section{Conclusions}
In the present paper we have presented the first results on variability and rotation of the G- and K-type  members of the  $\sim$ 500-Myr old open cluster NGC\,2099, which has been studied as a part of the RACE-OC project.
The photometric accuracy achieved in our observations has allowed us to explore presence of variability from F to middle K-type stars.
Among 1746 candidate cluster members, we identified 122 periodic variables (16 of which were already known periodic variables). Whereas the periodicity of F-type stars likely arises from the presence of pulsations, for 80 G- and K-type stars the detected periodicity arises from starspot  activity and represents the stars' rotation period.
We have first time  determined  the  distribution of rotation period for these active  cluster variables and found that rotation period of G and K stars have median value of P=5.9 and P=7.1 days, respectively. This mass-dependent difference tells us that, although K-type stars have reached the ZAMS later than G-type stars and, consequently, have started later experiencing  the rotation magnetic braking, the braking has been more
effective for K- than for G-type stars. The newly determined rotation periods of the NGC\,2099 candidate members are very important since they allow us to partly fill the gap in the empirical description of the angular momentum evolution in the age interval from 200  to 600 Myr.  A  comparison with rotation data from the younger M34 ($\sim$ 200 Myr)  and the older Hyades clusters ($\sim$ 625 Myr), shows that the NGC\,2099  members rotate faster than the older Hyades stars, with a clear mass dependency, as expected from angular momentum evolution models. The very interesting result is that G8-K5 members of NGC\,2099 have the same rotation period distribution of the younger  G8-K5 members of M34. These stars seem to have experienced no braking in the age range from 200 to 500 Myr. Finally, the new light curve amplitude data of NGC\,2099 give some further support to the suspect that the level of magnetic activity, in the photosphere at least, has some other not yet identified
 age dependence, which makes stars older than Pleiades
less and less active, independently on mass and rotation.

\begin{acknowledgements}
This work was supported  by the  Italian Ministero della Ricerca (MUR), Italian Ministero degli Affari Esteri (MAE) and Indian Department of Science and 
Technology (DST).
The extensive  use of  the SIMBAD  and ADS databases  operated   by  the  CDS  center,   
Strasbourg,  France,  is gratefully acknowledged. The work of S.-C. R was supported in part by KOSEF through
the Astrophysical Research Center for the Structure and Evolution
of the Cosmos (ARCSEC).
\end{acknowledgements}

\Online

\begin{deluxetable}{r r r r l}
\tablecolumns{12}
\tablecaption{\label{tab-photom} ID of periodic variables, RA and DEC coordinates (J2000), average V magnitude (V$_{\rm mean}$)$_0$ corrected for extinction and de-reddened (B$-$V)$_0$ color.}
\small
\tablewidth{0pt}
 \tablehead{
\colhead {ID}   & \colhead {RA} & \colhead {DEC}  & \colhead {(V$_{\rm mean}$)$_0$} & \colhead {(B$-$V)$_0$}\\   
  
\colhead {}   & \colhead {(hh:mm:ss)} & \colhead {(dd:mm:ss)}  & \colhead {(mag)}   & \colhead {(mag)} \\    
}
\startdata
   59  &  05:52:11.286  &  +32:34:11.06  & 13.281  &  0.209  \\
   80  &  05:52:21.216  &  +32:34:16.91  & 14.451  &  0.492  \\
   96  &  05:52:27.800  &  +32:33:38.69  & 13.941  &  0.369  \\
  263  &  05:52:12.768  &  +32:28:41.15  & 14.301  &  0.426  \\
  267  &  05:52:08.778  &  +32:28:33.38  & 13.801  &  0.350  \\
  290  &  05:51:59.524  &  +32:34:16.49  & 12.911  &  0.153  \\
  295  &  05:52:03.702  &  +32:35:14.21  & 14.741  &  0.317  \\
  320  &  05:52:14.955  &  +32:37:06.04  & 14.331  &  0.276  \\
  352  &  05:52:40.942  &  +32:32:01.70  & 13.651  &  0.325  \\
  370  &  05:52:33.520  &  +32:28:59.89  & 13.831  &  0.363  \\
  385  &  05:52:08.058  &  +32:27:34.82  & 13.961  &  0.502  \\
  414  &  05:51:47.281  &  +32:32:11.84  & 13.611  &  0.249  \\
  441  &  05:52:00.816  &  +32:36:48.67  & 13.881  &  0.431  \\
  483  &  05:52:50.944  &  +32:34:48.84  & 14.871  &  0.620  \\
  486  &  05:52:45.611  &  +32:33:38.36  & 12.361  &  0.118  \\
  491  &  05:52:46.210  &  +32:32:35.88  & 14.471  &  0.433  \\
  492  &  05:52:45.834  &  +32:32:20.36  & 14.181  &  0.369  \\
  493  &  05:52:46.479  &  +32:32:12.72  & 14.881  &  0.479  \\
  512  &  05:52:35.713  &  +32:26:38.04  & 13.781  &  0.262  \\
  548  &  05:51:47.557  &  +32:28:26.05  & 14.651  &  0.345  \\
  566  &  05:51:38.218  &  +32:31:47.94  & 14.551  &  0.133  \\
  570  &  05:51:40.520  &  +32:32:32.48  & 13.811  &  0.419  \\
  595  &  05:52:02.827  &  +32:40:09.13  & 15.111  &  0.172  \\
  643  &  05:52:53.438  &  +32:29:51.64  & 13.831  &  0.310  \\
  666  &  05:52:12.182  &  +32:24:34.15  & 13.841  &  0.691  \\
  761  &  05:53:01.748  &  +32:28:40.53  & 12.851  &  0.135  \\
  764  &  05:52:50.548  &  +32:26:27.94  & 13.741  &  0.345  \\
  786  &  05:52:00.961  &  +32:22:20.29  & 13.001  &  0.283  \\
  855  &  05:52:36.952  &  +32:43:18.78  & 13.771  &  0.084  \\
  857  &  05:52:43.801  &  +32:43:49.72  & 13.981  &  0.528  \\
  871  &  05:53:10.740  &  +32:33:47.22  & 13.200  &  0.312  \\ 
  916  &  05:51:30.381  &  +32:25:22.05  & 15.111  &  0.404  \\
  952  &  05:51:51.693  &  +32:43:49.99  & 12.821  &  1.008  \\
  969  &  05:53:06.085  &  +32:41:36.58  & 14.511  &  0.351  \\
 2112  &  05:51:56.717  &  +32:35:16.54  & 15.981  &  0.687  \\
 2257  &  05:52:26.549  &  +32:40:45.29  & 15.351  &  0.549  \\
 2312  &  05:52:18.846  &  +32:33:36.14  & 16.041  &  0.654  \\
 2318  &  05:52:22.711  &  +32:32:28.78  & 14.821  &  0.580  \\
 2321  &  05:52:18.530  &  +32:31:50.81  & 16.521  &  0.799  \\
 2330  &  05:52:13.920  &  +32:33:19.59  & 16.321  &  0.640  \\
 2385  &  05:52:27.231  &  +32:34:45.24  & 15.501  &  0.334  \\
 2395  &  05:52:31.316  &  +32:32:49.32  & 16.511  &  0.765  \\
 2408  &  05:52:24.266  &  +32:30:32.11  & 16.661  &  0.839  \\
 2409  &  05:52:22.439  &  +32:30:30.90  & 16.421  &  0.726  \\
 2443  &  05:52:26.522  &  +32:36:23.97  & 15.901  &  0.663  \\
 2444  &  05:52:27.104  &  +32:36:10.71  & 15.501  &  0.607  \\
 2456  &  05:52:33.022  &  +32:32:41.79  & 15.340  &  0.295  \\
 2471  &  05:52:28.149  &  +32:29:55.11  & 15.961  &  0.668  \\
 2472  &  05:52:27.495  &  +32:29:41.03  & 16.911  &  0.957  \\
 2473  &  05:52:23.246  &  +32:29:19.41  & 16.451  &  0.747  \\
 2485  &  05:52:01.813  &  +32:32:11.12  & 15.031  &  0.469  \\
 2506  &  05:52:19.780  &  +32:37:30.37  & 16.901  &  0.692  \\
 2509  &  05:52:20.404  &  +32:38:44.19  & 16.391  &  0.916  \\
 2530  &  05:52:31.672  &  +32:36:32.97  & 15.381  &  0.447  \\
 2549  &  05:52:37.672  &  +32:35:17.70  & 15.561  &  0.551  \\
 2557  &  05:52:38.088  &  +32:34:16.43  & 16.211  &  0.440  \\
 2562  &  05:52:44.972  &  +32:33:51.22  & 16.101  &  0.632  \\
 2595  &  05:52:42.188  &  +32:29:38.50  & 16.591  &  0.809  \\
 2597  &  05:52:39.476  &  +32:29:00.09  & 16.781  &  0.808  \\
 2610  &  05:52:32.822  &  +32:30:24.92  & 16.151  &  0.688  \\
 2618  &  05:52:33.751  &  +32:28:06.57  & 16.251  &  0.787  \\
 2635  &  05:52:15.993  &  +32:28:04.51  & 15.631  &  0.591  \\
 2639  &  05:52:14.653  &  +32:28:45.35  & 16.831  &  0.781  \\
 2646  &  05:52:12.320  &  +32:27:47.38  & 17.131  &  0.961  \\
 2650  &  05:52:08.712  &  +32:28:55.25  & 15.541  &  0.850  \\
 2665  &  05:51:58.835  &  +32:29:56.21  & 16.371  &  0.727  \\
 2667  &  05:52:00.312  &  +32:30:33.97  & 15.141  &  0.505  \\
 2676  &  05:51:58.275  &  +32:31:15.56  & 14.821  &  0.478  \\
 2681  &  05:51:57.034  &  +32:31:36.18  & 15.591  &  0.626  \\
 2687  &  05:51:51.187  &  +32:32:36.17  & 15.501  &  0.508  \\
 2701  &  05:51:52.245  &  +32:34:19.87  & 14.711  &  0.489  \\
 2728  &  05:52:03.061  &  +32:38:19.21  & 15.751  &  0.645  \\
 2742  &  05:52:17.219  &  +32:37:27.87  & 15.101  &  0.507  \\
 2743  &  05:52:18.470  &  +32:40:22.54  & 16.231  &  0.592  \\
 2763  &  05:52:26.055  &  +32:40:01.70  & 16.211  &  0.675  \\
 2832  &  05:52:53.511  &  +32:33:01.48  & 15.261  &  0.599  \\
 2835  &  05:53:00.720  &  +32:32:06.26  & 15.681  &  0.424  \\
 2843  &  05:52:53.330  &  +32:31:41.30  & 16.361  &  0.692  \\
 2857  &  05:52:44.894  &  +32:30:17.51  & 15.372  &  0.480  \\
 2859  &  05:52:46.090  &  +32:29:36.08  & 15.621  &  0.609  \\
 2862  &  05:52:50.636  &  +32:29:00.59  & 15.061  &  1.261  \\
 2874  &  05:52:53.400  &  +32:27:14.14  & 15.621  &  0.436  \\
 2895  &  05:52:38.385  &  +32:27:19.72  & 15.371  &  0.595  \\
 2947  &  05:52:19.351  &  +32:26:22.48  & 15.151  &  0.543  \\
 2948  &  05:52:22.001  &  +32:25:15.94  & 15.441  &  0.715  \\
 2968  &  05:52:14.556  &  +32:23:48.43  & 16.341  &  0.729  \\
 2971  &  05:52:12.781  &  +32:24:02.70  & 15.661  &  0.772  \\
 2973  &  05:52:11.646   & +32:25:17.97  & 16.056  &  0.149  \\
 2974  &  05:52:12.321  &  +32:26:10.25  & 15.651  &  0.584  \\
 2995  &  05:52:08.122  &  +32:26:40.53  & 15.731  &  0.563  \\
 2996  &  05:52:09.242  &  +32:26:41.95  & 16.431  &  0.733  \\
 2997  &  05:52:06.024  &  +32:27:08.38  & 16.101  &  0.703  \\
 3003  &  05:52:02.651  &  +32:24:54.32  & 16.311  &  0.705  \\
 3008  &  05:51:57.938  &  +32:25:06.26  & 16.291  &  0.623  \\
 3021  &  05:52:01.447  &  +32:26:46.04  & 15.021  &  0.628  \\
 3030  &  05:51:48.881  &  +32:26:26.38  & 16.561  &  0.714  \\
 3040  &  05:51:49.605  &  +32:28:13.18  & 15.051  &  0.477  \\
 3082  &  05:51:43.681  &  +32:30:40.53  & 16.641  &  0.761  \\
 3089  &  05:51:34.133  &  +32:30:40.44  & 15.081  &  0.511  \\
 3090  &  05:51:32.827  &  +32:31:45.45  & 15.811  &  0.313  \\
 3094  &  05:51:49.489  &  +32:31:38.83  & 16.041  &  0.663  \\
 3110  &  05:51:32.724  &  +32:32:26.20  & 15.541  &  0.533  \\
 3146  &  05:51:50.368  &  +32:35:13.53  & 15.131  &  0.477  \\
 3157  &  05:51:47.902  &  +32:36:01.59  & 15.531  &  0.578  \\
 3209  &  05:51:58.156  &  +32:39:46.40  & 15.661  &  0.602  \\
 3230  &  05:52:07.245  &  +32:39:14.15  & 16.441  &  0.713  \\
 3236  &  05:52:06.222  &  +32:41:25.42  & 16.261  &  0.854  \\
 3245  &  05:52:08.640  &  +32:41:18.84  & 16.591  &  0.779  \\
 3246  &  05:52:11.343  &  +32:41:44.50  & 16.652  &  0.524  \\
 3247  &  05:52:10.953  &  +32:41:19.94  & 16.141  &  0.596  \\
 3452  &  05:52:58.280  &  +32:38:44.74  & 14.521  &  0.708  \\
 3550  &  05:53:07.503  &  +32:30:58.67  & 16.125  &  0.271 \\
 3638  &  05:53:09.357  &  +32:26:23.33  & 16.081  &  0.818  \\
 3714  &  05:52:36.626  &  +32:23:01.65  & 16.301  &  0.738  \\
 3724  &  05:53:00.859  &  +32:24:51.65  & 15.411  &  0.604  \\
 3725  &  05:52:58.565  &  +32:23:26.78  & 16.351  &  0.561  \\
 3866  &  05:52:34.784  &  +32:28:25.90  & 15.281  &  0.624  \\
 3909  &  05:51:56.038  &  +32:33:25.03  & 16.901  &  0.541  \\
 4035  &  05:52:42.224  &  +32:27:31.65  & 15.121  &  0.562  \\
 4035  &  05:52:42.224  &  +32:27:31.65  & 15.121  &  0.562  \\
 4103  &  05:51:33.453  &  +32:39:28.59  & 16.281  &  0.807  \\
 4115  &  05:52:08.437  &  +32:32:15.76  & 16.221  & -2.206  \\
 4134  &  05:52:46.232  &  +32:25:34.30  & 16.691  &  0.823  \\
 4181  &  05:52:39.365  &  +32:36:29.83  & 16.951  &  0.566  \\
 4326  &  05:52:24.415  &  +32:24:44.93  & 17.561  &  1.268  \\
 4402  &  05:52:47.663  &  +32:39:33.20  & 17.199  &  0.588  \\
 4423  &  05:51:47.965  &  +32:36:51.93  & 17.281  &  0.904  \\
 4481  &  05:52:48.453  &  +32:36:42.43  & 17.281  &  0.840  \\
 4576  &  05:53:08.465  &  +32:33:47.47  & 18.001  &  0.755  \\
 4654  &  05:53:03.640  &  +32:32:12.83  & 17.877  &  0.691  \\
 4708  &  05:52:41.016  &  +32:24:24.60  & 17.359  &  0.859  \\
 4822  &  05:53:04.053  &  +32:29:41.74  & 18.211  &  1.022  \\
 5082  &  05:52:15.809  &  +32:43:36.20  & 15.821  &  0.712  \\
 5893  &  05:51:34.735  &  +32:29:06.72  & 20.066  &  0.411  \\
 5906  &  05:52:44.159  &  +32:28:52.05  & 19.749  &  0.646  \\
 6040  &  05:52:49.442  &  +32:26:17.52  & 18.301  &  1.176  \\
 6462  &  05:52:57.457  &  +32:41:15.44  & 17.221  &  0.709  \\
 7941  &  05:51:29.856  &  +32:24:19.35  & 16.896  &  0.594  \\

\hline
\enddata
\end{deluxetable}

\clearpage

\begin{deluxetable}{r r r r r l l l l c c c}
\tablecolumns{12}
\tablecaption{\label{tab-photom} Star's ID, period detected by Scargle (P$_{\rm Scargle}$) and by CLEAN (P$_{\rm CLEAN}$), normalized peak power (P$_{\rm N}$), adopted rotation period (P) and its uncertainty, average V magnitude ($<$V$>$), standard deviation ($\sigma_{\rm TOT}$), achieved photometric accuracy, light curve amplitude ($\Delta$V), number of observations,  note about the membership (cm is candidate member). }
\small
\tablewidth{0pt}
 \tablehead{
\colhead {ID}   & \colhead {P$_{\rm Scargle}$} & \colhead {P$_{\rm CLEAN}$} & \colhead {P$_{\rm N}$ } & \colhead {P$\pm\Delta$P}   & \colhead {V$_{\rm mean}$} & \colhead {$\sigma_{\rm TOT}$} & \colhead { accuracy} &
 \colhead {$\Delta$V}      & \colhead { \# obs. }       & \colhead {note}   \\   
  
 \colhead {}   & \colhead {(d)} & \colhead {(d)} & \colhead {} & \colhead {(d)}   & \colhead {(mag)} & \colhead {(mag)} & \colhead { (mag)} &
 \colhead {(mag)}      & \colhead {}       & \colhead {}  \\    
}
\startdata
\multicolumn{11}{c}{\bf F-type periodic candidate members}\\
star00080.dat  &  5.433  &  0.000  &  15.31  &  5.433 $\pm$ 0.095   &  15.13  & 0.0065 &  0.0031 &  0.0185 &  362  &   cm  \\
star00096.dat  &  2.170  &  0.000  &  15.80  &  2.170 $\pm$ 0.014   &  14.61  & 0.0118 &  0.0041 &  0.0338 &  362  &   cm  \\
star00263.dat  &  5.393  &  5.400  &  15.42  &  5.393 $\pm$ 0.048   &  14.98  & 0.0115 &  0.0033 &  0.0318 &  362  &   cm  \\
star00290.dat  &  0.339  &  0.337  &  16.49  &  0.339 $\pm$ 0.003   &  13.59  & 0.0066 &  0.0030 &  0.0176 &  358  &   cm  \\
star00295.dat  &  0.106  &  0.119  &  93.56  &  0.119 $\pm$ 0.000   &  15.42  & 0.0378 &  0.0100 &  0.1080 &  362  &   $\delta$ Scuti (KV4)  \\
star00320.dat  &  5.354  &  0.840  &  15.44  &  5.354 $\pm$ 0.003   &  15.01  & 0.0077 &  0.0034 &  0.0210 &  362  &   cm  \\
star00370.dat  &  0.558  &  0.000  &  27.99  &  0.558 $\pm$ 0.003   &  14.51  & 0.0205 &  0.0039 &  0.0605 &  362  &   cm  \\
star00385.dat  &  1.911  &  1.899  &  35.21  &  1.911 $\pm$ 0.005   &  14.64  & 0.0222 &  0.0046 &  0.0565 &  362  &   cm  \\
star00441.dat  &  0.944  &  0.952  & 140.31  &  1.888 $\pm$ 0.003   &  14.56  & 0.0391 &  0.0100 &  0.1147 &  362  &  EW (KV11)  \\
star00491.dat  &  1.732  &  1.732  &  21.00  &  1.732 $\pm$ 0.063   &  15.15  & 0.0078 &  0.0029 &  0.0215 &  362  &   cm  \\
star00492.dat  &  5.433  &  5.391  &  19.29  &  5.433 $\pm$ 0.069   &  14.86  & 0.0075 &  0.0030 &  0.0209 &  362  &   cm  \\
star00493.dat  &  9.533  & 10.000  &  17.96  &  9.533 $\pm$ 0.279   &  15.56  & 0.0069 &  0.0034 &  0.0203 &  362  &   cm  \\
star00512.dat  &  0.574  &  0.578  &  24.87  &  0.578 $\pm$ 0.003   &  14.46  & 0.0186 &  0.0028 &  0.0481 &  362  &   cm  \\
star00548.dat  &  0.556  &  0.602  &  16.02  &  0.556 $\pm$ 0.003   &  15.33  & 0.0103 &  0.0037 &  0.0283 &  362  &   cm  \\
star00643.dat  &  3.762  &  0.787  &  20.73  &  3.762 $\pm$ 0.010   &  14.51  & 0.0113 &  0.0028 &  0.0323 &  362  &   cm  \\
star00916.dat  &  3.602  &  0.000  &  18.67  &  3.602 $\pm$ 0.034   &  15.79  & 0.0165 &  0.0040 &  0.0338 &  362  &   cm  \\
star02456.dat  &  0.211  &  0.212  &  79.10  &  0.422 $\pm$ 0.003   &  16.02  & 0.1025 &  0.0100 &  0.2835 &  362  &   EW (V3)  \\
star02485.dat  &  1.712  &  1.706  &  16.37  &  1.712 $\pm$ 0.008   &  15.71  & 0.0076 &  0.0033 &  0.0176 &  362  &   cm  \\
star02596.dat  & 11.834  &  0.000  &  21.67  & 11.834 $\pm$ 0.250   &  16.18  & 0.0076 &  0.0039 &  0.0209 &  362  &   cm  \\
star02667.dat  &  3.821  &  3.650  &  29.70  &  3.821 $\pm$ 0.023   &  15.82  & 0.0089 &  0.0041 &  0.0233 &  362  &   cm  \\
star02676.dat  &  5.393  &  5.391  &  15.73  &  5.393 $\pm$ 0.084   &  15.50  & 0.0070 &  0.0030 &  0.0178 &  362  &   cm  \\
star02687.dat  &  0.099  &  0.110  &  84.38  &  0.220 $\pm$ 0.003   &  16.18  & 0.1520 &  0.0100 &  0.4304 &  362  &   $\delta$ Scuti (V6) \\
star02701.dat  &  3.981  &  0.000  &  17.20  &  3.981 $\pm$ 0.040   &  15.39  & 0.0050 &  0.0023 &  0.0143 &  362  &   cm  \\
star02742.dat  &  5.135  &  5.155  &  20.30  &  5.155 $\pm$ 0.065   &  15.77  & 0.0092 &  0.0043 &  0.0270 &  362  &   cm  \\
star02835.dat  &  5.473  &  5.391  &  15.46  &  5.391 $\pm$ 0.077   &  16.36  & 0.0160 &  0.0066 &  0.0399 &  362  &   cm  \\
star02874.dat  & 10.209  & 10.000  &  16.20  & 10.209 $\pm$ 0.317   &  16.29  & 0.0076 &  0.0040 &  0.0222 &  362  &   cm  \\
star03031.dat  &  7.065  &  0.000  &  21.78  &  7.065 $\pm$ 0.081   &  15.84  & 0.0099 &  0.0038 &  0.0235 &  362  &   cm  \\
star03040.dat  &  2.687  &  2.674  &  23.95  &  2.687 $\pm$ 0.017   &  15.73  & 0.0093 &  0.0035 &  0.0259 &  362  &   cm  \\
star03089.dat  &  5.990  &  0.000  &  22.62  &  5.990 $\pm$ 0.058   &  15.76  & 0.0065 &  0.0036 &  0.0196 &  362  &   cm  \\
star03146.dat  &  3.742  &  0.000  &  21.34  &  3.742 $\pm$ 0.031   &  15.81  & 0.0095 &  0.0038 &  0.0275 &  362  &   cm  \\
star03246.dat  &  0.042  &  0.044  &  19.63  &  0.044 $\pm$ 0.003   &  17.33  & 0.0404 &  0.0100 &  0.1128 &  362  &   $\delta$ Scuti (KV10)  \\
star05893.dat  &  0.112  &  0.127  &  18.30  &  0.224 $\pm$ 0.003   &  20.83  & 0.3004 &  0.0100 &  0.8412 &  345  &   EW (KV17) \\

\multicolumn{11}{c}{\bf G-type periodic candidate members}\\
star00427.dat  &  5.672  &  5.500  &  19.01  &  5.500 $\pm$ 0.200   &  15.29  & 0.0068 &  0.0031 &  0.0196 &  362  &   cm  \\
star00483.dat  &  5.413  &  5.391  &  40.48  &  5.413 $\pm$ 0.032   &  15.55  & 0.0130 &  0.0028 &  0.0404 &  362  &   cm  \\
star02112.dat  &  5.732  &  5.900  &  17.76  &  5.900 $\pm$ 0.101   &  16.66  & 0.0150 &  0.0054 &  0.0295 &  362  &   cm  \\
star02257.dat  &  5.294  &  5.155  &  18.30  &  5.294 $\pm$ 0.074   &  16.03  & 0.0121 &  0.0050 &  0.0331 &  362  &   cm  \\
star02312.dat  &  9.234  &  9.217  &  17.30  &  9.217 $\pm$ 0.258   &  16.72  & 0.0150 &  0.0068 &  0.0444 &  362  &   cm  \\
star02318.dat  &  5.373  &  5.391  &  17.47  &  5.373 $\pm$ 0.079   &  15.50  & 0.0137 &  0.0040 &  0.0390 &  362  &   cm  \\
star02330.dat  &  5.851  &  5.900  &  21.99  &  5.900 $\pm$ 0.076   &  17.00  & 0.0134 &  0.0068 &  0.0364 &  362  &   cm  \\
star02409.dat  &  7.125  &  7.117  &  19.55  &  7.125 $\pm$ 0.152   &  17.10  & 0.0149 &  0.0068 &  0.0401 &  362  &   cm  \\
star02443.dat  &  6.806  &  6.711  &  24.20  &  6.806 $\pm$ 0.096   &  16.58  & 0.0105 &  0.0048 &  0.0292 &  362  &   cm  \\
star02444.dat  &  6.229  &  6.211  &  20.16  &  6.229 $\pm$ 0.097   &  16.18  & 0.0085 &  0.0038 &  0.0241 &  362  &   cm  \\
star02506.dat  &  8.557  &  8.264  &  18.61  &  8.557 $\pm$ 0.399   &  17.58  & 0.0168 &  0.0093 &  0.0444 &  362  &   cm  \\
star02549.dat  &  1.792  &  1.786  &  17.28  &  1.786 $\pm$ 0.009   &  16.24  & 0.0072 &  0.0037 &  0.0189 &  362  &   cm  \\
star02562.dat  &  1.752  &  1.758  &  17.20  &  1.758 $\pm$ 0.008   &  16.78  & 0.0105 &  0.0051 &  0.0300 &  362  &   cm  \\
star02610.dat  &  8.000  &  7.752  &  29.43  &  7.752 $\pm$ 0.133   &  16.83  & 0.0144 &  0.0063 &  0.0386 &  362  &   cm  \\
star02635.dat  &  5.553  &  0.000  &  29.85  &  5.553 $\pm$ 0.053   &  16.31  & 0.0115 &  0.0047 &  0.0299 &  362  &   cm  \\
star02665.dat  &  7.423  &  7.576  &  18.25  &  7.423 $\pm$ 0.134   &  17.05  & 0.0118 &  0.0062 &  0.0306 &  362  &   cm  \\
star02681.dat  &  5.553  &  5.618  &  23.51  &  5.618 $\pm$ 0.066   &  16.27  & 0.0111 &  0.0042 &  0.0276 &  362  &   cm  \\
star02703.dat  &  8.737  &  0.000  &  21.42  &  8.737 $\pm$ 0.307   &  16.10  & 0.0072 &  0.0034 &  0.0193 &  362  &   cm  \\
star02728.dat  &  6.567  &  0.000  &  22.47  &  6.567 $\pm$ 0.095   &  16.43  & 0.0124 &  0.0062 &  0.0367 &  362  &   cm  \\
star02743.dat  &  1.133  &  0.000  &  36.69  &  1.155 $\pm$ 0.003   &  16.92  & 0.0264 &  0.0075 &  0.0740 &  362  &   cm  \\
star02763.dat  &  1.314  &  1.316  &  17.72  &  1.314 $\pm$ 0.004   &  16.89  & 0.0164 &  0.0067 &  0.0457 &  362  &   cm  \\
star02832.dat  &  0.279  &  0.279  &  51.70  &  0.558 $\pm$ 0.003   &  15.95  & 0.0899 &  0.0100 &  0.2720 &  362  &   EW (V4)  \\
star02843.dat  &  4.876  &  4.938  &  17.61  &  4.876 $\pm$ 0.067   &  17.04  & 0.0132 &  0.0061 &  0.0373 &  362  &   cm  \\
star02847.dat  &  1.274  &  0.000  &  18.16  &  1.274 $\pm$ 0.004   &  15.62  & 0.0065 &  0.0029 &  0.0181 &  362  &   cm  \\
star02859.dat  &  6.329  &  6.211  &  26.34  &  6.329 $\pm$ 0.072   &  16.30  & 0.0081 &  0.0037 &  0.0230 &  362  &   cm  \\
star02895.dat  &  1.786  &  0.000  &  20.97  &  1.786 $\pm$ 0.004   &  16.05  & 0.0076 &  0.0036 &  0.0195 &  362  &   cm  \\
star02947.dat  &  3.702  &  3.984  &  20.64  &  3.984 $\pm$ 0.030   &  15.83  & 0.0052 &  0.0033 &  0.0153 &  362  &   cm  \\
star02948.dat  &  5.274  &  5.155  &  21.02  &  5.274 $\pm$ 0.067   &  16.12  & 0.0077 &  0.0039 &  0.0217 &  362  &   cm  \\
star02968.dat  &  6.349  &  6.211  &  22.83  &  6.211 $\pm$ 0.102   &  17.02  & 0.0149 &  0.0067 &  0.0434 &  362  &   cm  \\
star02974.dat  &  1.811  &  1.813  &  27.00  &  1.813 $\pm$ 0.007   &  16.33  & 0.0086 &  0.0046 &  0.0251 &  362  &   cm  \\
star02995.dat  &  0.493  &  0.000  &  22.30  &  0.986 $\pm$ 0.003   &  16.41  & 0.0315 &  0.0100 &  0.0882 &  362  &  EA (KV12)  \\
star02996.dat  &  6.110  &  6.211  &  18.56  &  6.211 $\pm$ 0.087   &  17.11  & 0.0112 &  0.0074 &  0.0277 &  362  &   cm  \\
star02997.dat  &  7.045  &  7.117  &  35.98  &  7.117 $\pm$ 0.045   &  16.78  & 0.0143 &  0.0052 &  0.0407 &  362  &   cm  \\
star03003.dat  &  6.906  &  0.000  &  27.28  &  6.906 $\pm$ 0.063   &  16.99  & 0.0137 &  0.0059 &  0.0394 &  362  &   cm  \\
star03008.dat  &  2.110  &  2.105  &  30.42  &  2.105 $\pm$ 0.008   &  16.97  & 0.0188 &  0.0054 &  0.0524 &  362  &   cm  \\
star03021.dat  & 10.110  & 10.000  &  18.42  & 10.000 $\pm$ 0.288   &  15.70  & 0.0403 &  0.0089 &  0.0975 &  362  &   cm  \\
star03030.dat  &  1.951  &  1.963  &  23.38  &  1.963 $\pm$ 0.008   &  17.24  & 0.0116 &  0.0064 &  0.0303 &  362  &   cm  \\
star03094.dat  &  3.224  &  0.000  &  46.64  &  3.224 $\pm$ 0.009   &  16.72  & 0.0442 &  0.0087 &  0.1196 &  362  &   cm  \\
star03110.dat  &  2.090  &  2.105  &  25.89  &  2.105 $\pm$ 0.009   &  16.22  & 0.0110 &  0.0042 &  0.0297 &  362  &   cm  \\
star03157.dat  &  5.771  &  5.900  &  33.64  &  5.900 $\pm$ 0.042   &  16.21  & 0.0110 &  0.0050 &  0.0329 &  362  &   cm  \\
star03208.dat  &  0.876  &  0.000  &  15.62  &  0.876 $\pm$ 0.002   &  16.97  & 0.0124 &  0.0075 &  0.0353 &  362  &   cm  \\
star03209.dat  &  5.771  &  5.900  &  21.75  &  5.900 $\pm$ 0.086   &  16.34  & 0.0155 &  0.0058 &  0.0434 &  362  &   cm  \\
star03230.dat  &  8.339  &  8.032  &  16.24  &  8.339 $\pm$ 0.332   &  17.12  & 0.0148 &  0.0086 &  0.0408 &  362  &   cm  \\
star03247.dat  & 12.139  & 11.834  &  18.62  & 11.834 $\pm$ 0.383   &  16.82  & 0.0136 &  0.0071 &  0.0392 &  362  &   cm  \\
star03294.dat  &  7.344  &  7.117  &  22.84  &  7.344 $\pm$ 0.128   &  16.62  & 0.0245 &  0.0109 &  0.0618 &  362  &   cm  \\
star03512.dat  &  6.508  &  0.000  &  17.07  &  6.508 $\pm$ 0.112   &  17.18  & 0.0135 &  0.0088 &  0.0331 &  362  &   cm  \\
star03617.dat  &  6.309  &  0.000  &  23.54  &  6.309 $\pm$ 0.080   &  16.62  & 0.0083 &  0.0051 &  0.0236 &  362  &   cm  \\
star03713.dat  &  3.642  &  0.000  &  20.24  &  3.642 $\pm$ 0.100   &  16.92  & 0.0158 &  0.0072 &  0.0437 &  362  &   cm  \\
star03714.dat  &  5.900  &  0.000  &  22.15  &  5.900 $\pm$ 0.030   &  16.98  & 0.0140 &  0.0071 &  0.0384 &  362  &   cm  \\
star03724.dat  &  0.279  &  0.279  &   0.00  &  0.558 $\pm$ 0.003   &  16.09  & 0.1361 &  0.0100 &  0.3652 &  361  &   RRc (V5)  \\
star03725.dat  &  6.110  &  0.000  &  15.51  &  6.110 $\pm$ 0.088   &  17.03  & 0.0159 &  0.0083 &  0.0421 &  362  &   cm  \\
star03866.dat  &  5.175  &  5.155  &  23.09  &  5.155 $\pm$ 0.064   &  15.96  & 0.0138 &  0.0047 &  0.0404 &  362  &   cm  \\
star04035.dat  &  4.040  &  0.000  &  16.30  &  4.040 $\pm$ 0.048   &  15.80  & 0.0094 &  0.0040 &  0.0251 &  361  &   cm  \\
star04181.dat  &  0.179  &  0.179  &  87.35  &  0.358 $\pm$ 0.003   &  17.63  & 0.1571 &  0.0100 &  0.4294 &  362  &   EW (V7)  \\
star04654.dat  &  0.177  &  0.178  &  16.00  &  0.354 $\pm$ 0.003   &  18.55  & 0.0850 &  0.0100 &  0.2372 &  362  &  EW (KV15)  \\
star04708.dat  &  0.167  &  0.168  &  35.90  &  0.335 $\pm$ 0.003   &  18.05  & 0.0671 &  0.0100 &  0.2024 &  361  &   EW (KV14)\\
star05082.dat  &  5.990  &  5.900  &  31.76  &  5.990 $\pm$ 0.056   &  16.50  & 0.0171 &  0.0052 &  0.0424 &  362  &   cm  \\
star05906.dat  &  0.148  &  0.147  &  23.60  &  0.295 $\pm$ 0.003   &  20.45  & 0.2087 &  0.0100 &  0.5837 &  357  &   EW (KV16)  \\
star06462.dat  &  3.722  &  0.000  &  20.73  &  3.722 $\pm$ 0.032   &  17.90  & 0.0457 &  0.0129 &  0.1280 &  362  &   cm  \\
star07941.dat  &  0.169  &  0.145  &  14.73  &  0.290 $\pm$ 0.003   &  17.65  & 0.0538 &  0.0100 &  0.1458 &  362  &  EW (KV13)  \\

\multicolumn{11}{c}{\bf K-type periodic candidate members}\\
star02321.dat  & 11.881  & 11.834  &  19.13  & 11.881 $\pm$ 0.325   &  17.20  & 0.0152 &  0.0080 &  0.0425 &  362  &   cm  \\
star02395.dat  &  3.992  &  0.000  &  21.68  &  3.992 $\pm$ 0.020   &  17.19  & 0.0186 &  0.0069 &  0.0540 &  362  &   cm  \\
star02408.dat  &  7.224  &  0.000  &  27.42  &  7.224 $\pm$ 0.071   &  17.34  & 0.0157 &  0.0085 &  0.0430 &  362  &   cm  \\
star02472.dat  &  1.473  &  0.000  &  34.77  &  1.473 $\pm$ 0.004   &  17.59  & 0.0233 &  0.0081 &  0.0666 &  362  &   cm  \\
star02473.dat  &  6.766  &  0.000  &  24.30  &  6.766 $\pm$ 0.076   &  17.13  & 0.0116 &  0.0072 &  0.0329 &  362  &   cm  \\
star02509.dat  & 14.468  & 13.158  &  15.82  & 14.468 $\pm$ 0.720   &  17.07  & 0.0127 &  0.0071 &  0.0354 &  362  &   cm  \\
star02541.dat  &  7.145  &  0.000  &  24.40  &  7.145 $\pm$ 0.116   &  17.53  & 0.0188 &  0.0087 &  0.0518 &  362  &   cm  \\
star02595.dat  &  8.737  &  0.000  &  21.84  &  8.737 $\pm$ 0.333   &  17.27  & 0.0159 &  0.0068 &  0.0438 &  362  &   cm  \\
star02597.dat  &  6.289  &  6.211  &  16.83  &  6.211 $\pm$ 0.094   &  17.46  & 0.0157 &  0.0071 &  0.0434 &  362  &   cm  \\
star02618.dat  &  7.117  &  0.000  &  18.67  &  7.117 $\pm$ 0.200   &  16.93  & 0.0116 &  0.0062 &  0.0330 &  362  &   cm  \\
star02639.dat  &  7.543  &  7.576  &  19.30  &  7.576 $\pm$ 0.147   &  17.51  & 0.0126 &  0.0066 &  0.0352 &  362  &   cm  \\
star02646.dat  &  7.204  &  0.000  &  16.12  &  7.204 $\pm$ 0.161   &  17.81  & 0.0180 &  0.0100 &  0.0518 &  362  &   cm  \\
star02734.dat  &  7.443  &  7.576  &  15.91  &  7.576 $\pm$ 0.152   &  17.08  & 0.0136 &  0.0067 &  0.0358 &  362  &   cm  \\
star02890.dat  &  8.239  &  8.065  &  24.48  &  8.239 $\pm$ 0.233   &  17.63  & 0.0195 &  0.0087 &  0.0510 &  362  &   cm  \\
star02971.dat  & 19.980  &  0.000  &  17.85  & 19.980 $\pm$ 0.964   &  16.34  & 0.0088 &  0.0039 &  0.0233 &  362  &   cm  \\
star03082.dat  &  6.806  &  6.711  &  28.90  &  6.806 $\pm$ 0.061   &  17.32  & 0.0142 &  0.0075 &  0.0432 &  362  &   cm  \\
star03236.dat  &  8.657  &  8.097  &  21.70  &  8.657 $\pm$ 0.463   &  16.94  & 0.0148 &  0.0068 &  0.0422 &  362  &   cm  \\
star03245.dat  &  5.391  &  0.000  &  27.45  &  5.391 $\pm$ 0.020   &  17.28  & 0.0215 &  0.0086 &  0.0577 &  362  &   cm  \\
star03582.dat  &  4.876  &  4.938  &  24.20  &  4.876 $\pm$ 0.058   &  17.28  & 0.0179 &  0.0076 &  0.0509 &  362  &   cm  \\
star03638.dat  &  6.189  &  0.000  &  16.88  &  6.189 $\pm$ 0.092   &  16.76  & 0.0151 &  0.0075 &  0.0415 &  362  &   cm  \\
star03702.dat  &  7.125  &  7.117  &  16.69  &  7.125 $\pm$ 0.141   &  16.97  & 0.0145 &  0.0056 &  0.0299 &  362  &   cm  \\
star03712.dat  &  7.145  &  7.117  &  20.74  &  7.145 $\pm$ 0.125   &  16.85  & 0.0156 &  0.0077 &  0.0450 &  362  &   cm  \\
star04103.dat  &  5.393  &  0.855  &  39.13  &  5.393 $\pm$ 0.021   &  16.96  & 0.0275 &  0.0095 &  0.0777 &  362  &   cm  \\
star04134.dat  &  5.680  &  5.680  &  14.66  &  5.680 $\pm$ 0.063   &  17.37  & 0.0131 &  0.0079 &  0.0343 &  362  &   cm  \\
star04423.dat  &  7.901  &  8.032  &  22.91  &  7.901 $\pm$ 0.155   &  17.96  & 0.0267 &  0.0136 &  0.0718 &  362  &   cm  \\
star04481.dat  &  9.055  &  0.000  &  30.99  &  9.055 $\pm$ 0.165   &  17.96  & 0.0252 &  0.0118 &  0.0704 &  362  &   cm  \\
star04822.dat  &  2.428  &  2.445  &  19.94  &  2.428 $\pm$ 0.013   &  18.89  & 0.0488 &  0.0248 &  0.1343 &  362  &   cm  \\
star06040.dat  &  1.035  &  0.000  &  36.90  &  1.035 $\pm$ 0.003   &  18.98  & 0.0813 &  0.0295 &  0.2338 &  362  &   cm  \\
\hline
\enddata
\end{deluxetable}
\clearpage
\begin{figure*}
\begin{minipage}{18cm}
\centering
\includegraphics[scale = 0.8, trim = 0 0 0 0, clip, angle=0]{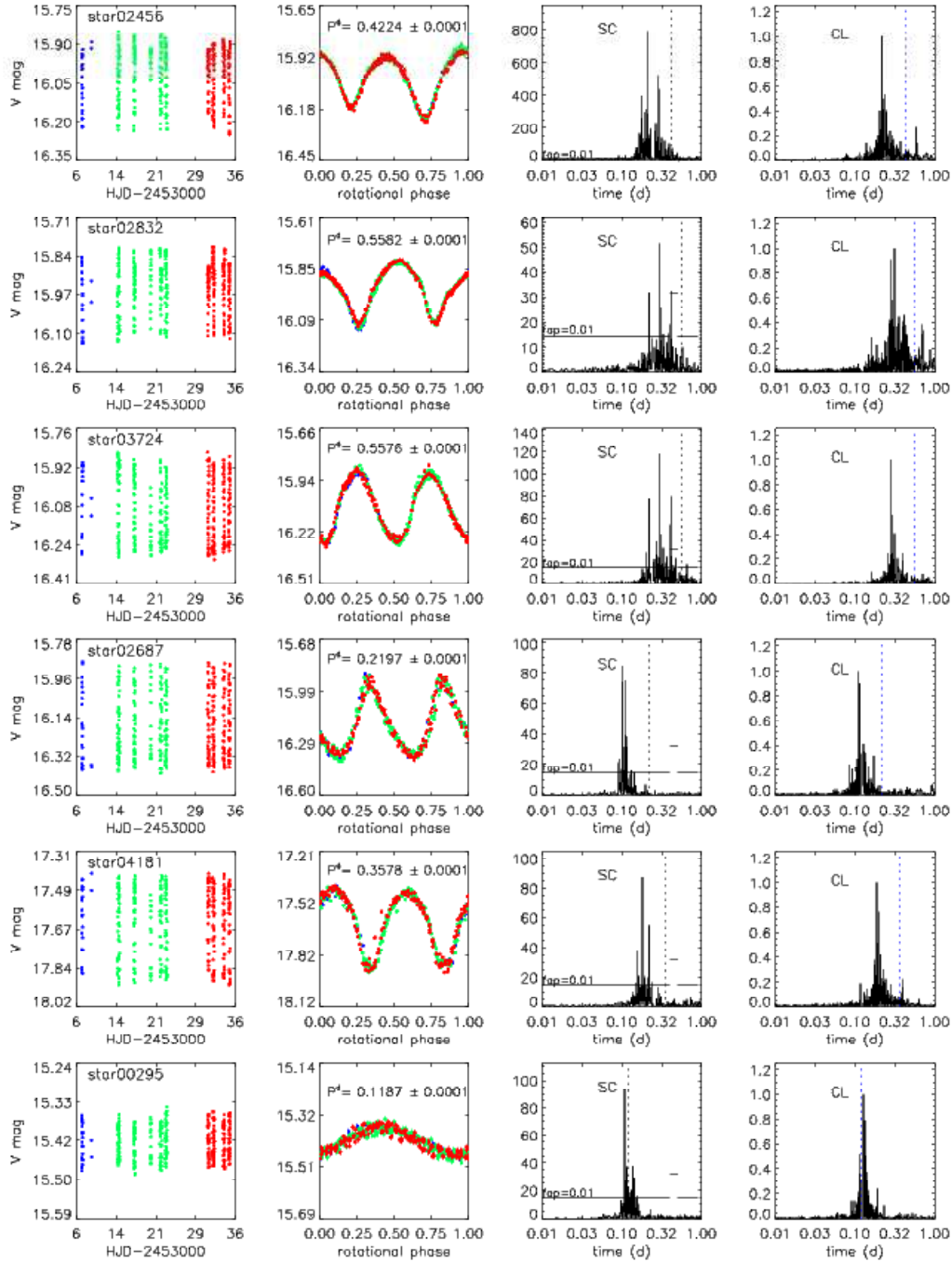}
\caption{\label{known_star1}Results of Fourier analysis of  periodic variables in M37 previously discovered by Kiss et al. (2001) and Kang et al. (2007) surveys. From left panel: V-band time series; phased light curve; Scargle periodogram and CLEAN periodogram. In case of eclipsing binaries the periodicity detected by either Scargle or CLEAN falls at exactly half the orbital period (vertical dashed line). See Sect. 3.3 for a detailed description.}
\end{minipage}
\end{figure*}

\begin{figure*}
\begin{minipage}{18cm}
\centering
\includegraphics[scale = 0.8, trim = 0 0 0 0, clip, angle=0]{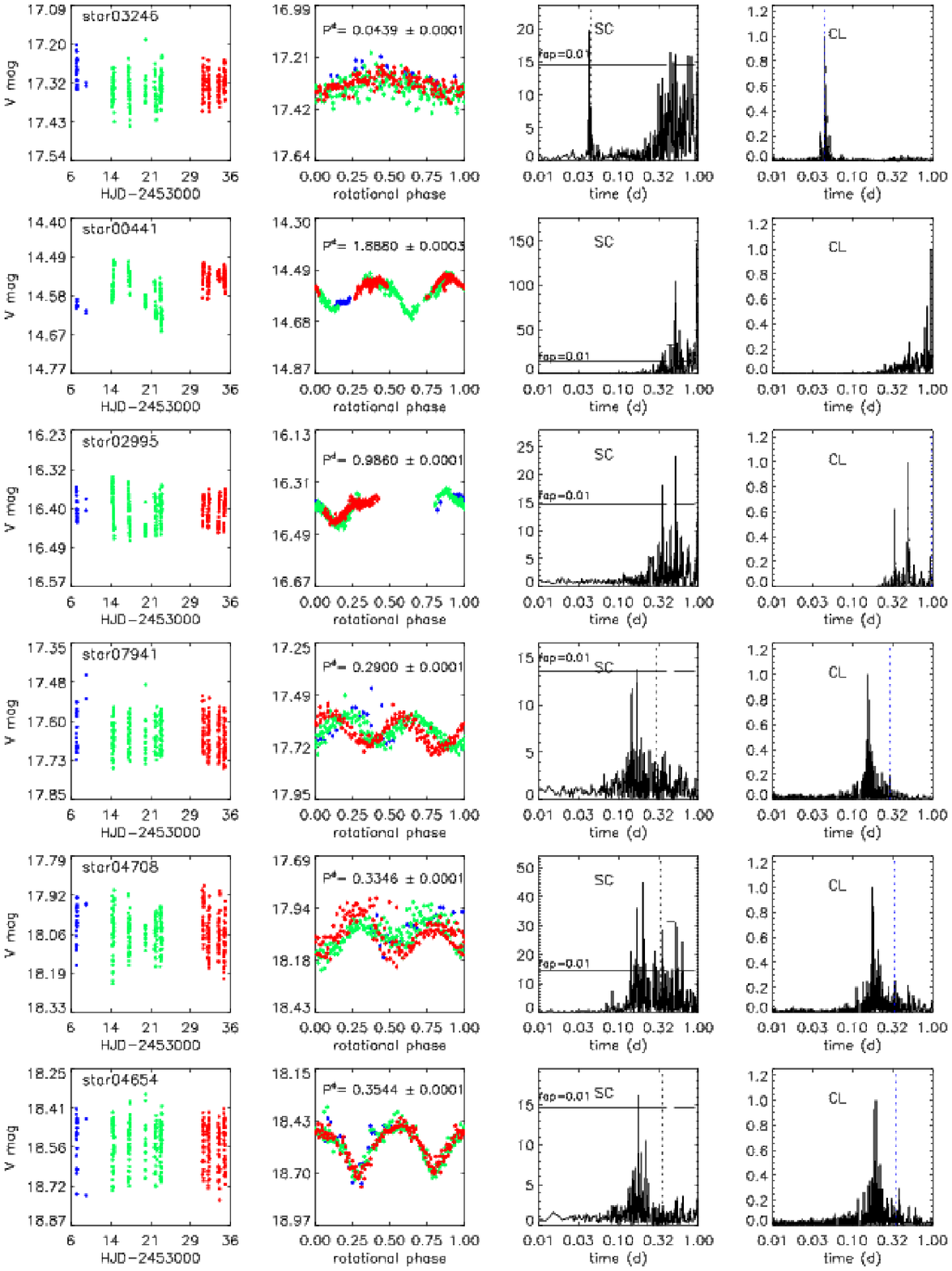}
\caption{\label{known_star1} As in Fig.\ref{known_star1}. }
\end{minipage}
\end{figure*}

\begin{figure*}
\begin{minipage}{18cm}
\centering
\includegraphics[scale = 0.8, trim = 0 100 0 0, clip, angle=270]{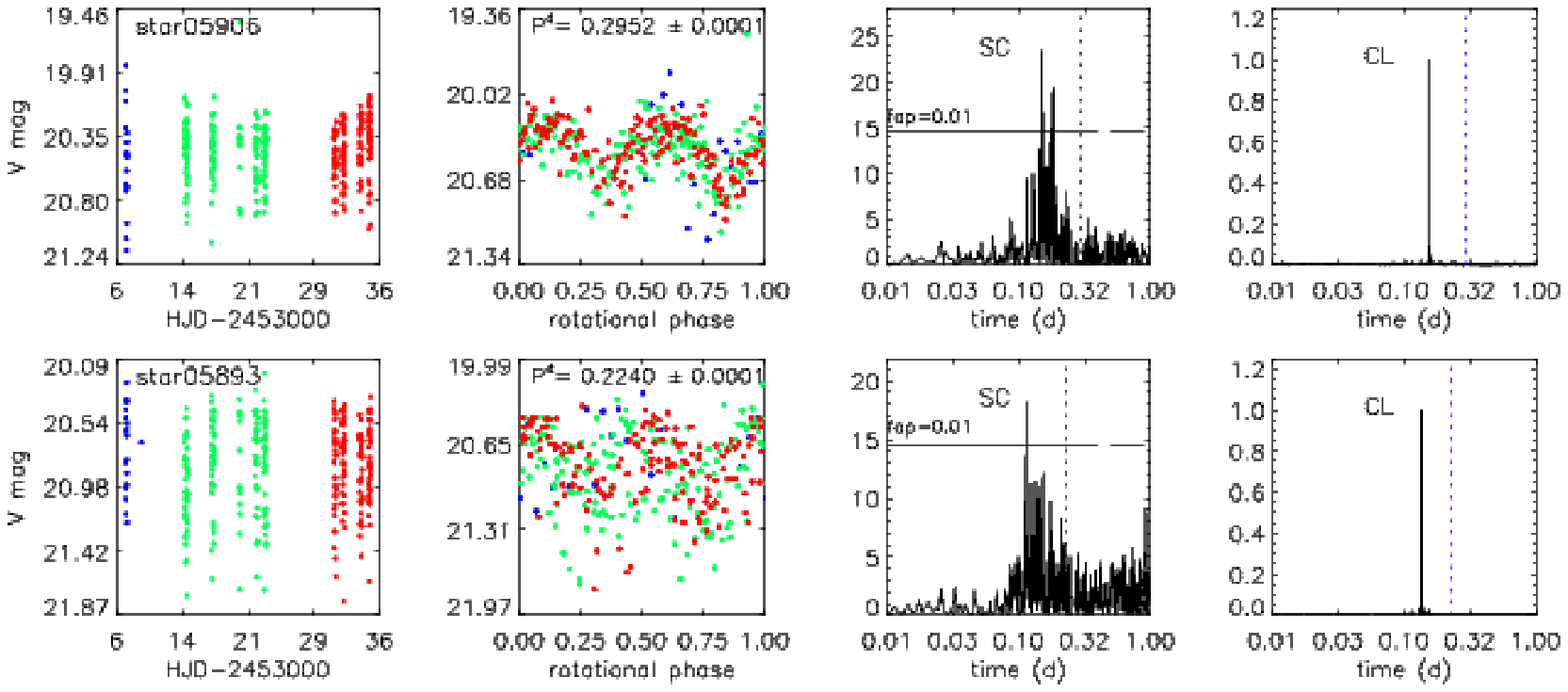}
\caption{\label{known_star1} As in Fig.\ref{known_star1}. }
\end{minipage}
\end{figure*}

\begin{figure*}
\begin{minipage}{18cm}
\centering
\includegraphics[scale = 0.94, trim = 0 0 0 100, clip, angle=0]{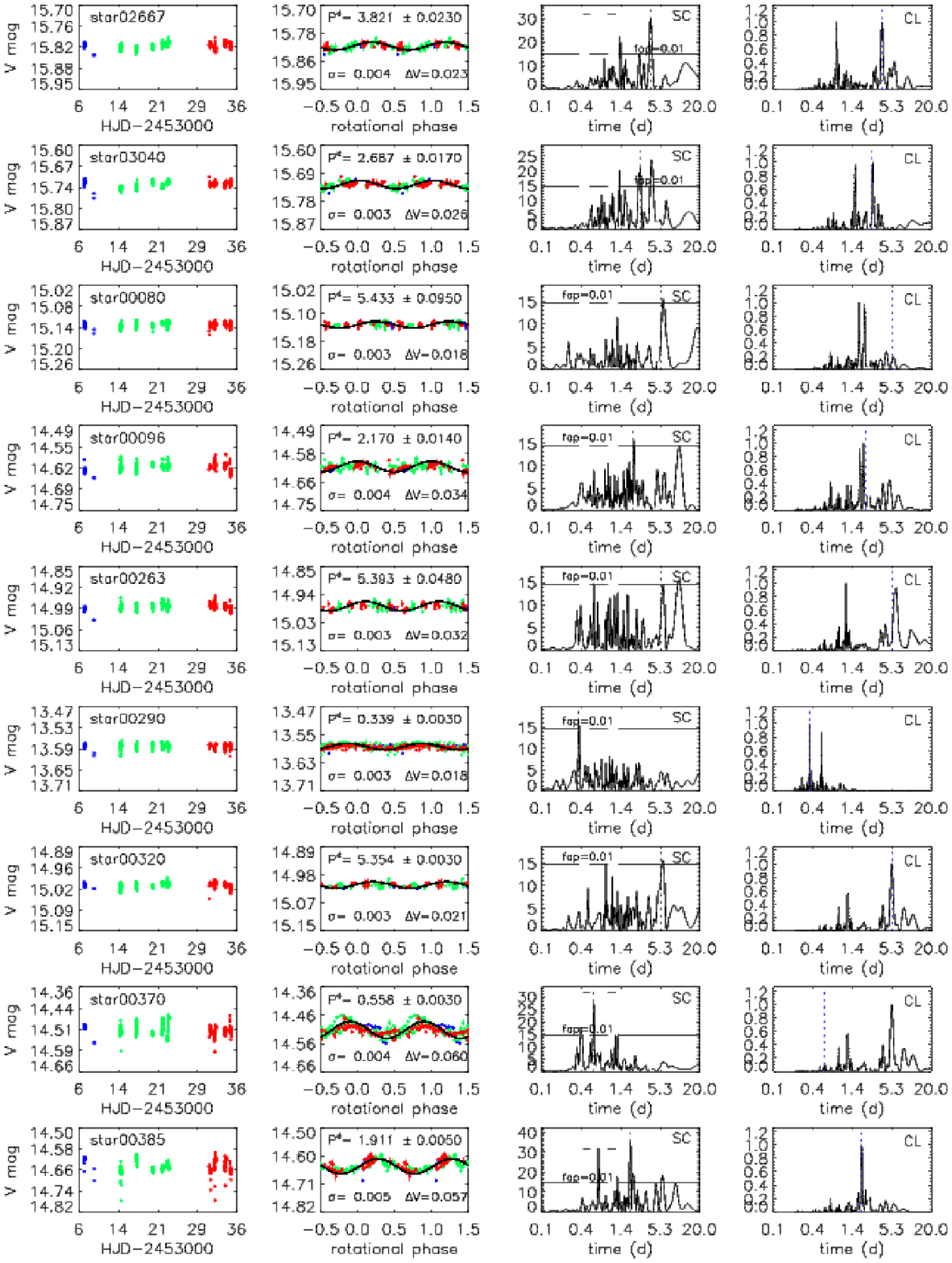}
\caption{\label{Fstar1} F-type periodic candidate cluster members. From left panel: V-band time series; phased light curve; Scargle and CLEAN periodogram. See Sect. 3.3 for a detailed description.}
\end{minipage}
\end{figure*}

\begin{figure*}
\begin{minipage}{18cm}
\centering
\includegraphics[scale = 0.94, trim = 0 0 0 100, clip, angle=0]{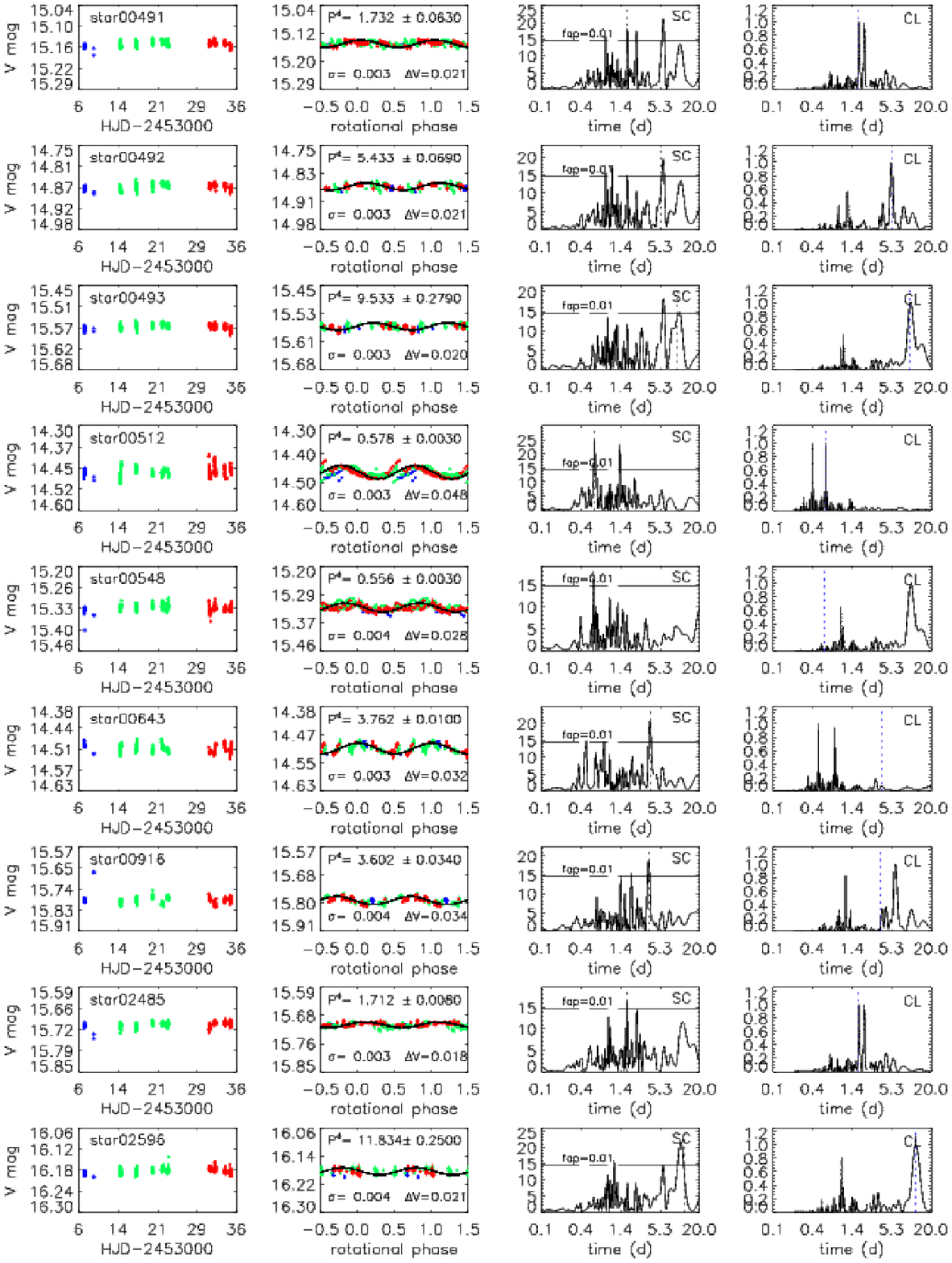}
\caption{\label{Fstar2} As in Fig.\ref{Fstar1}. }
\end{minipage}
\end{figure*}

\begin{figure*}
\begin{minipage}{18cm}
\centering
\includegraphics[scale = 0.94, trim = 0 100 0 100, clip, angle=0]{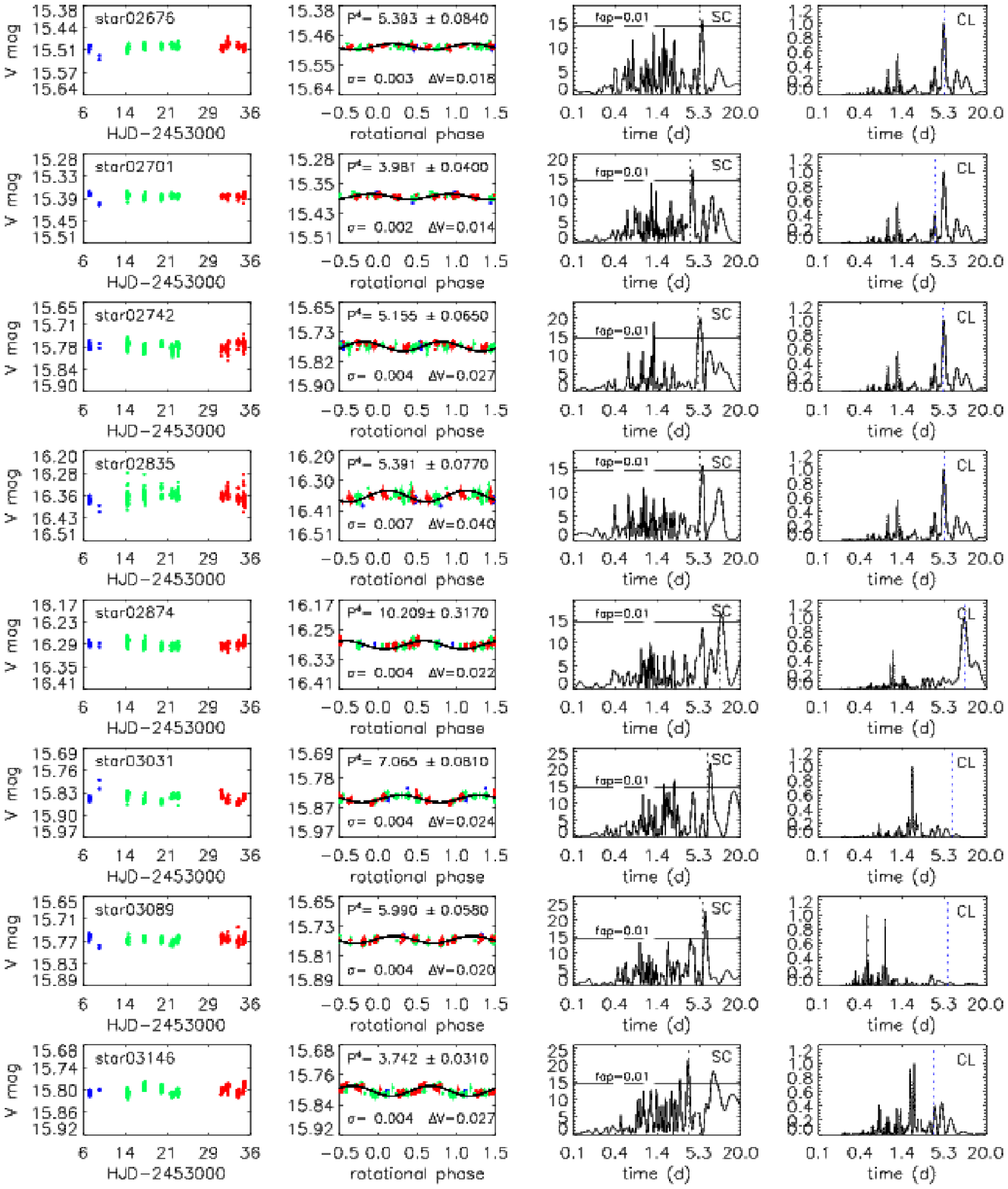}
\caption{\label{Fstar3} As in Fig.\ref{Fstar1}. }
\end{minipage}
\end{figure*}

\begin{figure*}
\begin{minipage}{18cm}
\centering
\includegraphics[scale = 0.94, trim = 0 0 0 100, clip, angle=0]{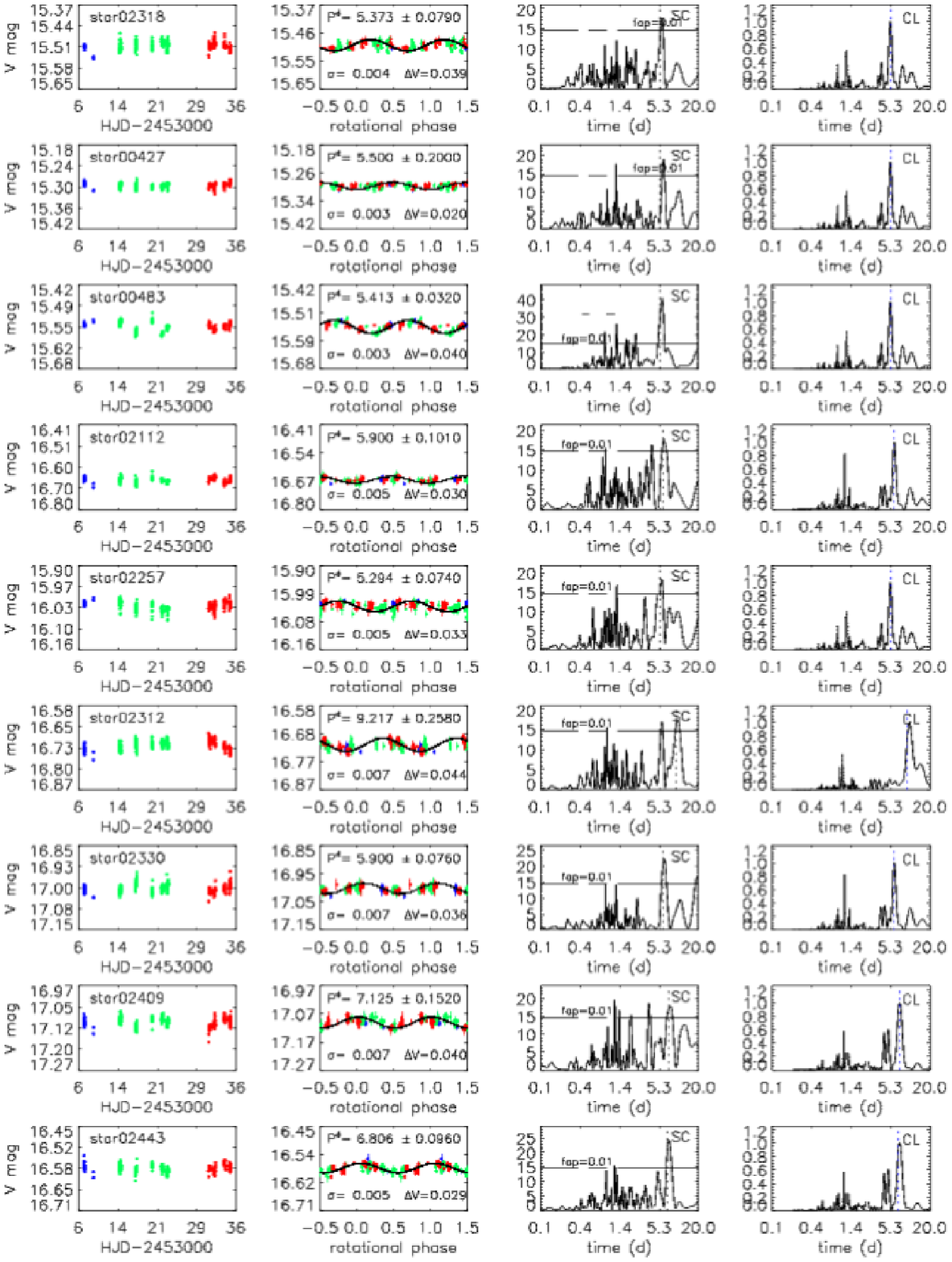}
\caption{\label{Gstar1} G-type periodic candidate cluster members. From left panel: V-band time series; phased light curve; Scargle and CLEAN periodogram. See Sect. 3.3 for a detailed description.}
\end{minipage}
\end{figure*}

\begin{figure*}
\begin{minipage}{18cm}
\centering
\includegraphics[scale = 0.94, trim = 0 0 0 100, clip, angle=0]{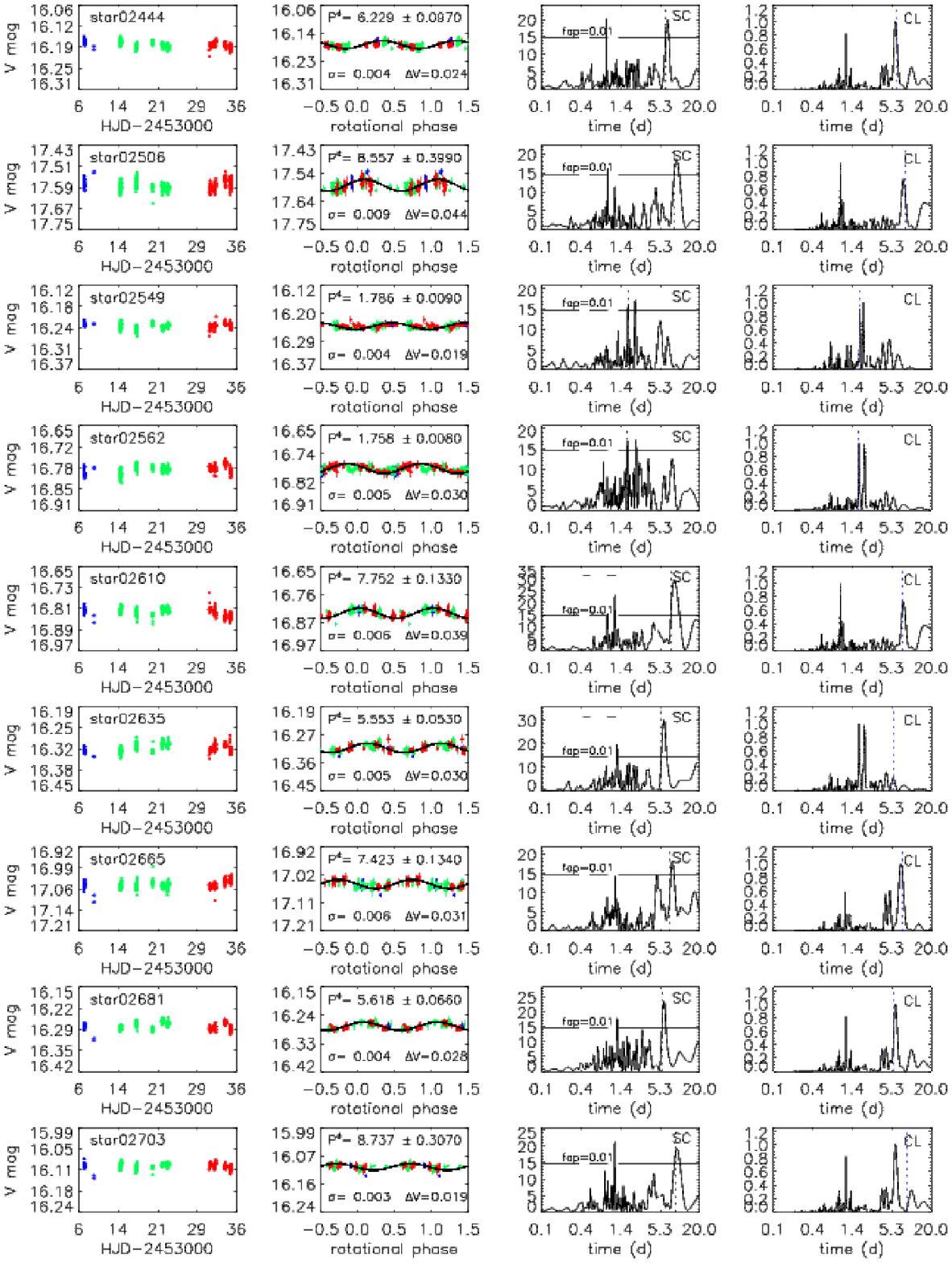}
\caption{\label{Gstar2} As in Fig.\ref{Gstar1}. }
\end{minipage}
\end{figure*}

\begin{figure*}
\begin{minipage}{18cm}
\centering
\includegraphics[scale = 0.94, trim = 0 0 0 100, clip, angle=0]{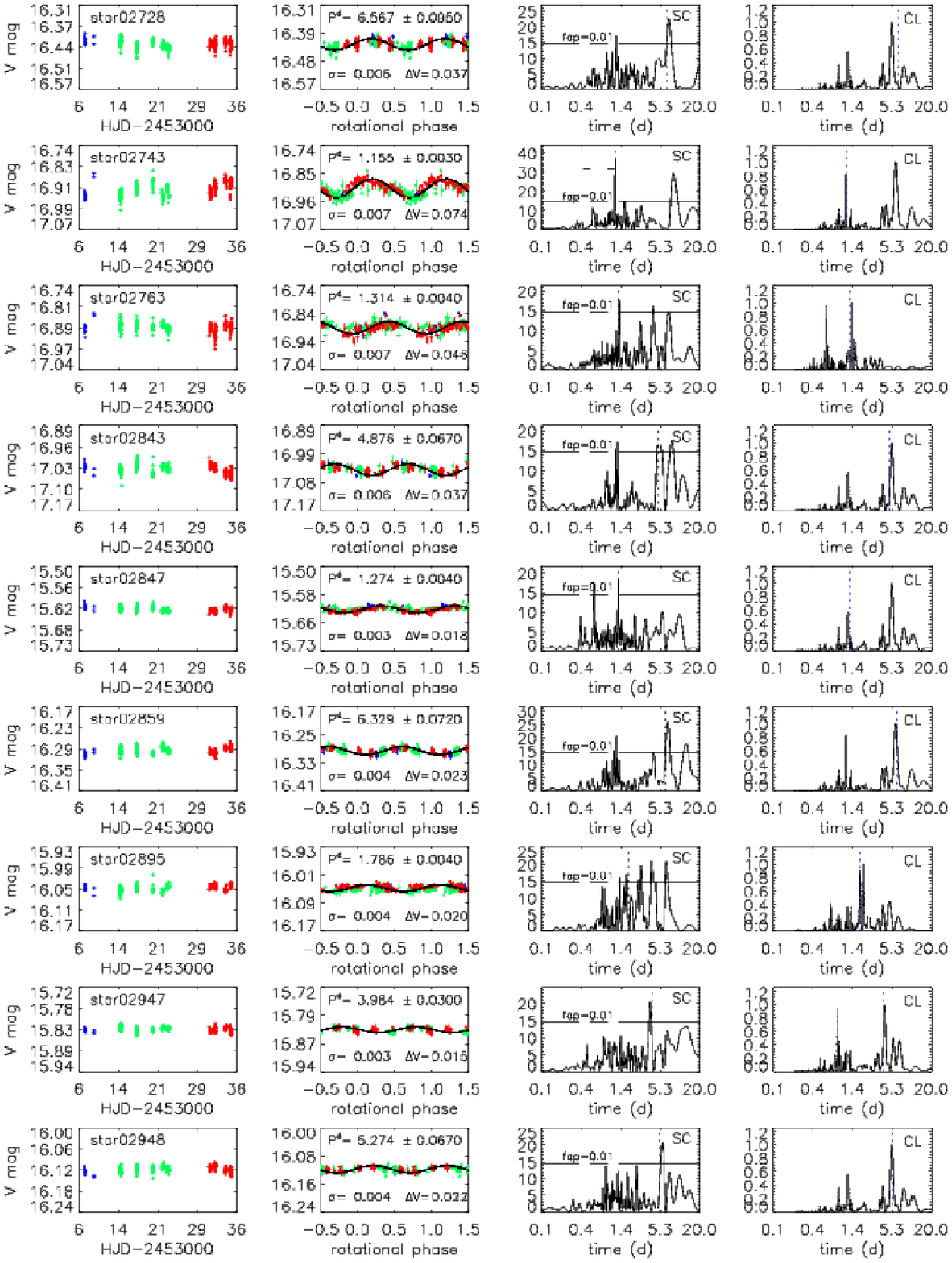}
\caption{\label{Gstar3} As in Fig.\ref{Gstar1}. }
\end{minipage}
\end{figure*}

\begin{figure*}
\begin{minipage}{18cm}
\centering
\includegraphics[scale = 0.94, trim = 0 0 0 100, clip, angle=0]{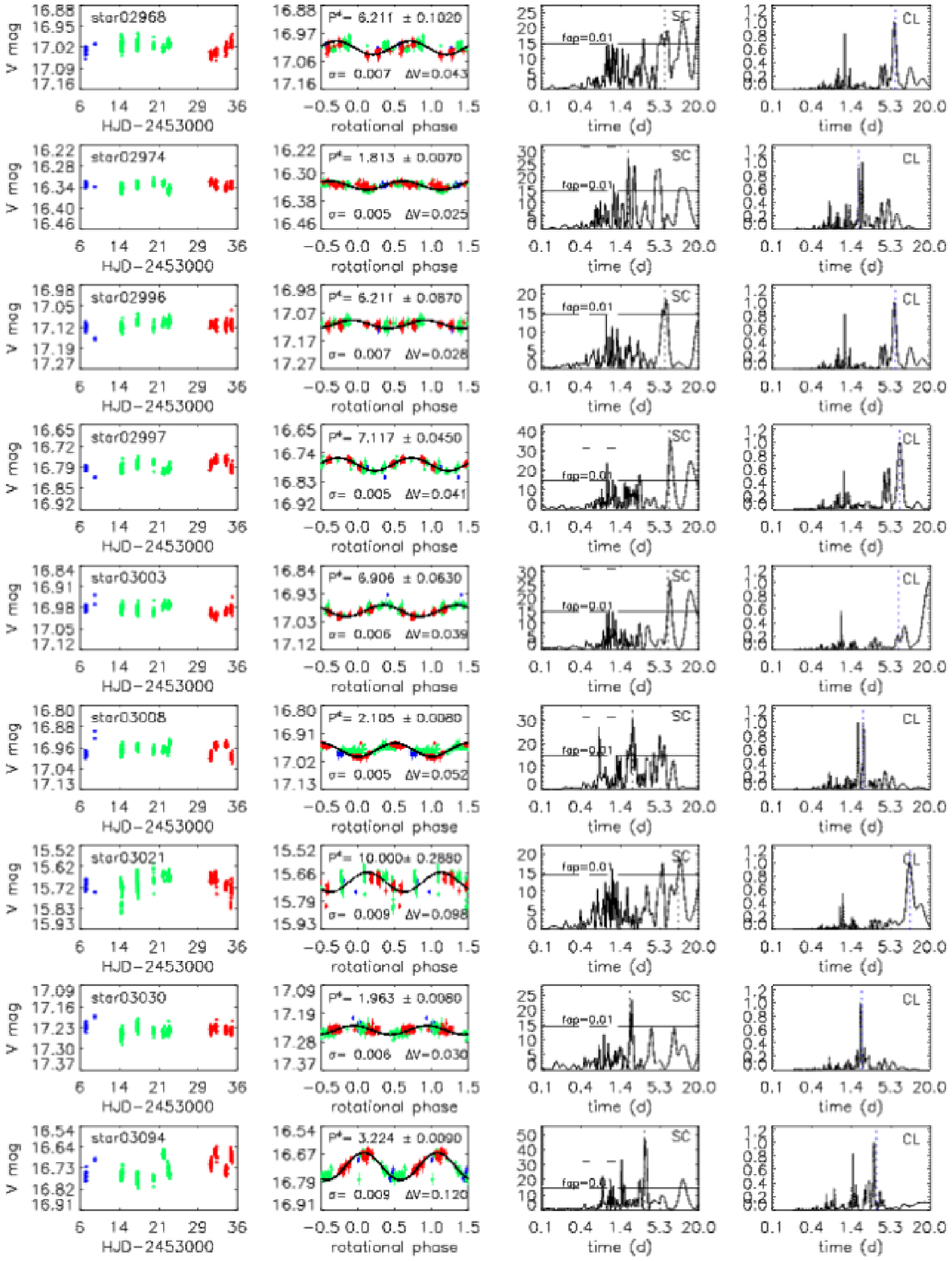}
\caption{\label{Gstar4} As in Fig.\ref{Gstar1}. }
\end{minipage}
\end{figure*}

\begin{figure*}
\begin{minipage}{18cm}
\centering
\includegraphics[scale = 0.94, trim = 0 0 0 100, clip, angle=0]{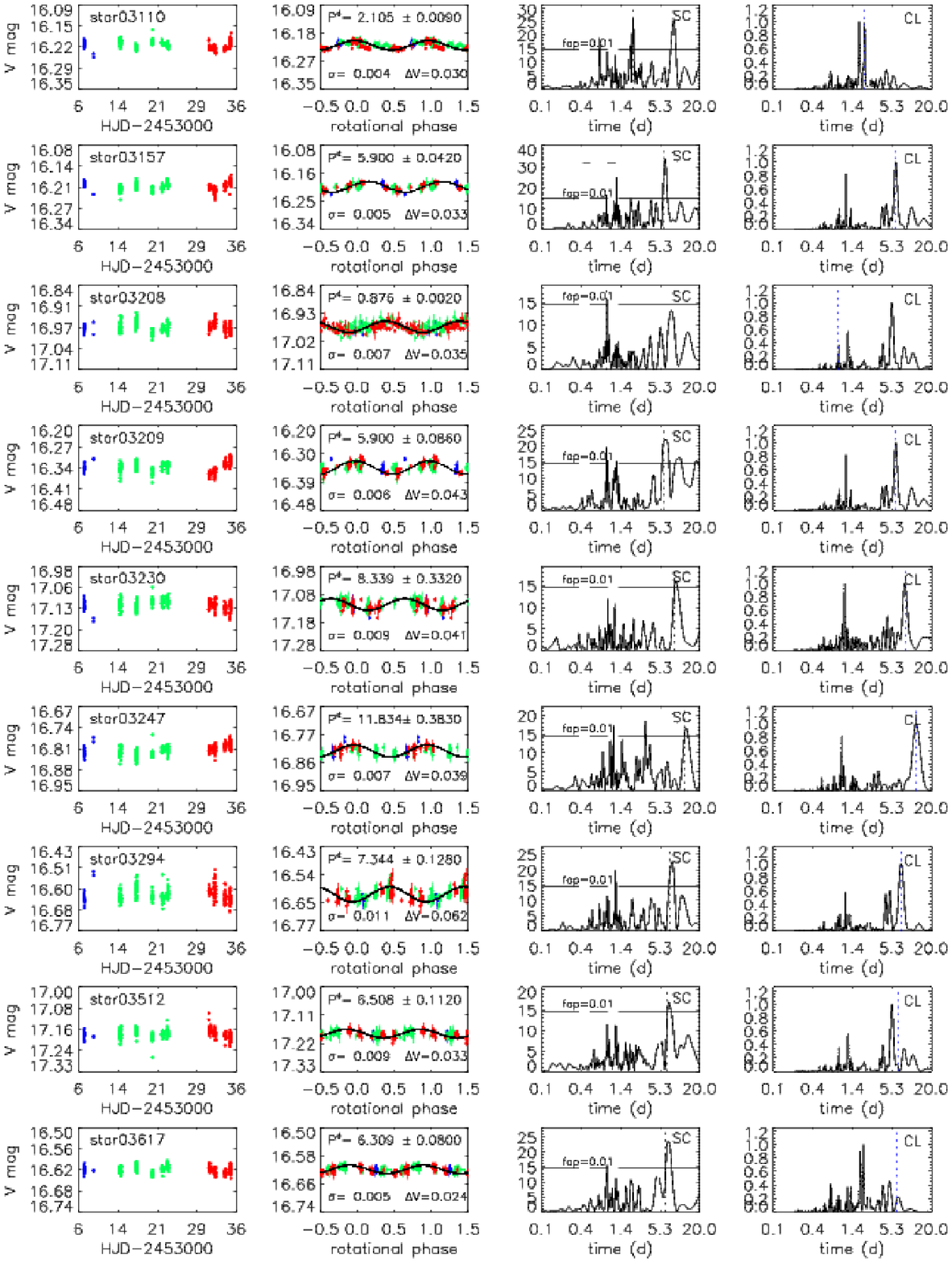}
\caption{\label{Gstar5} As in Fig.\ref{Gstar1}. }
\end{minipage}
\end{figure*}

\begin{figure*}
\begin{minipage}{18cm}
\centering
\includegraphics[scale = 0.94, trim = 0 200 0 0, clip, angle=0]{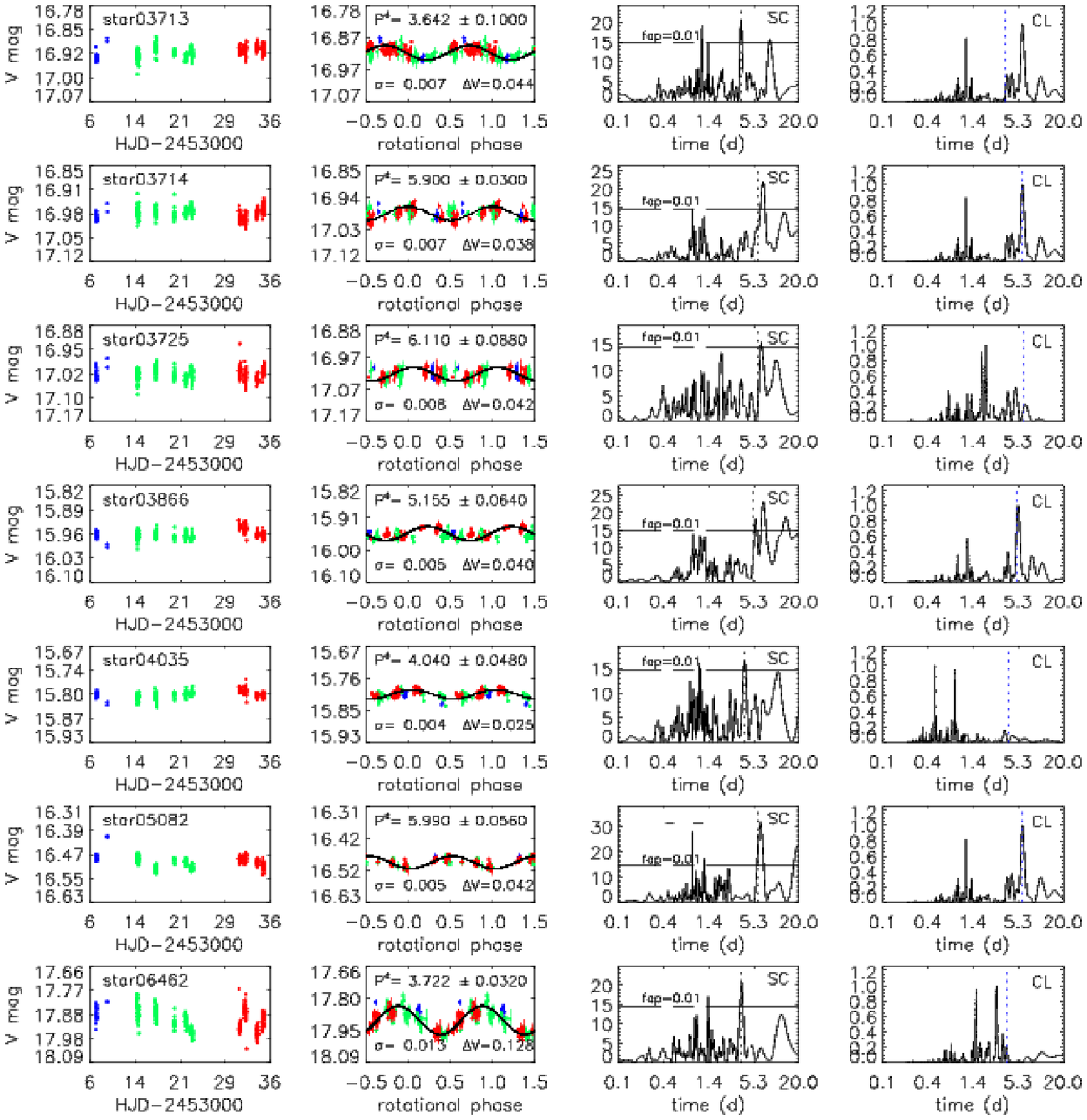}
\caption{\label{Gstar6} As in Fig.\ref{Gstar1}. }
\end{minipage}
\end{figure*}

\begin{figure*}
\begin{minipage}{18cm}
\centering
\includegraphics[scale = 0.94, trim = 0 0 0 100, clip, angle=0]{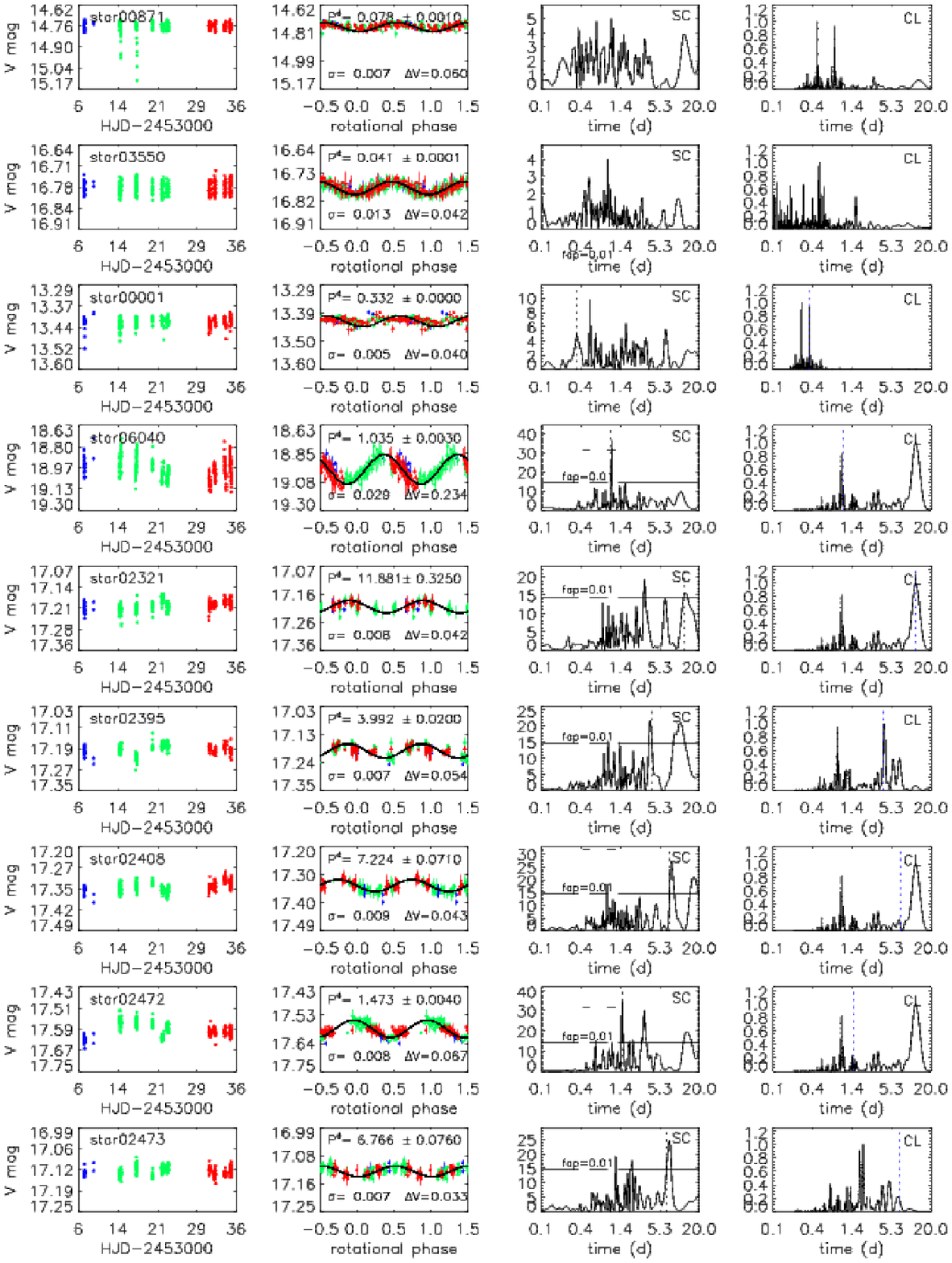}
\caption{\label{Kstar1} K-type periodic candidate cluster members. From left panel: V-band time series; phased light curve; Scargle and CLEAN periodogram. See Sect. 3.3 for a detailed description.}
\end{minipage}
\end{figure*}

\begin{figure*}
\begin{minipage}{18cm}
\centering
\includegraphics[scale = 0.94, trim = 0 0 0 100, clip, angle=0]{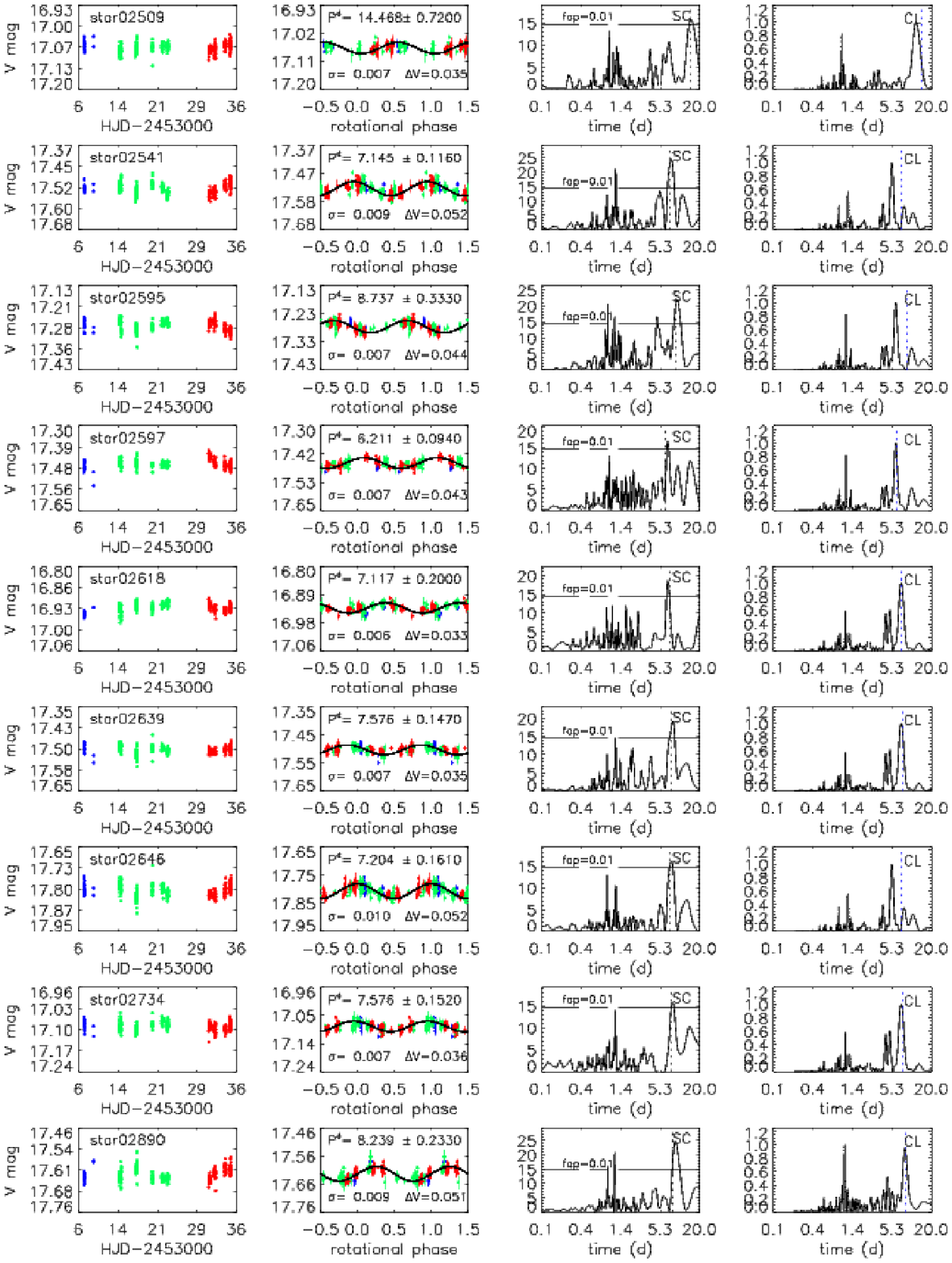}
\caption{\label{Kstar2} As in Fig.\ref{Kstar1}. }
\end{minipage}
\end{figure*}

\begin{figure*}
\begin{minipage}{18cm}
\centering
\includegraphics[scale = 0.94, trim = 0 0 0 100, clip, angle=0]{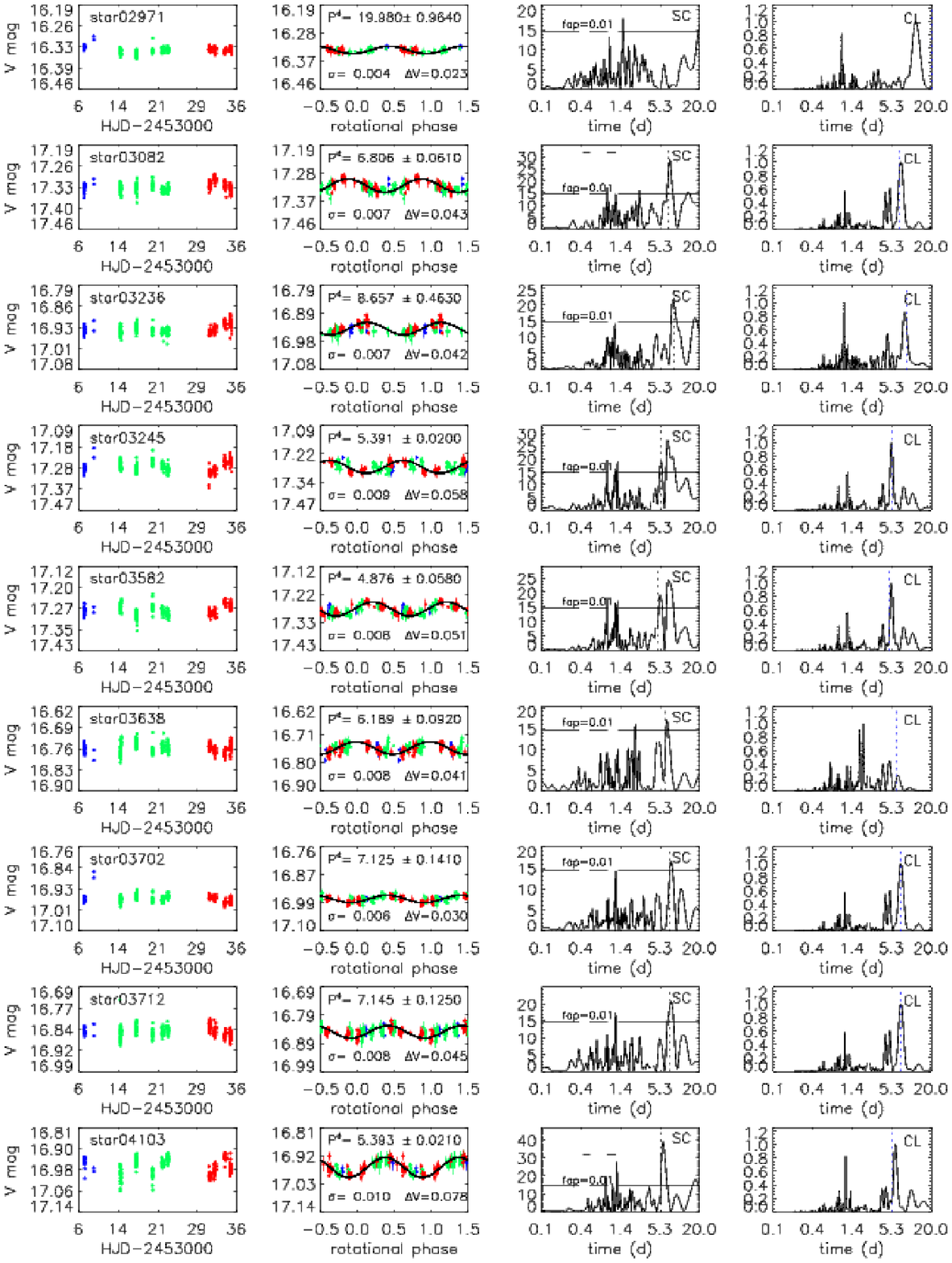}
\caption{\label{Kstar3} As in Fig.\ref{Kstar1}. }
\end{minipage}
\end{figure*}

\begin{figure*}
\begin{minipage}{18cm}
\centering
\includegraphics[scale = 0.94, trim = 0 00 0 00, clip, angle=270]{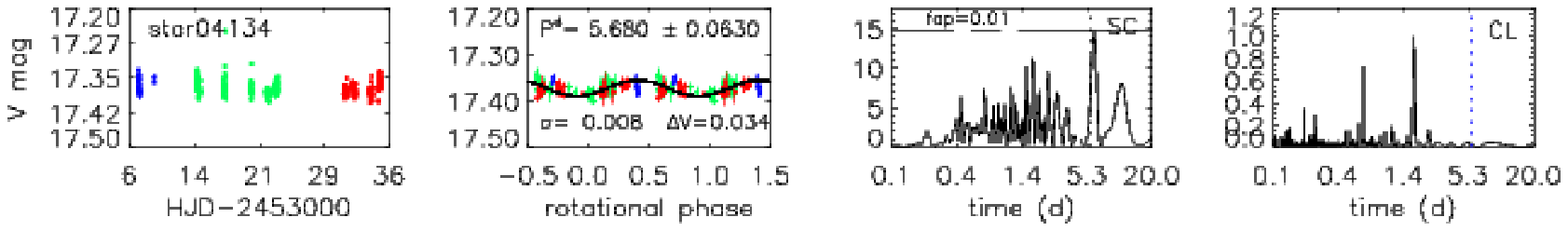}
\caption{\label{Kstar4} As in Fig.\ref{Kstar1}. }
\end{minipage}
\end{figure*}

\end{document}